%% file: main.tex
\documentclass[sigconf]{acmart}

\usepackage{stfloats}
\usepackage{multirow}
\usepackage{url}
\usepackage{makecell}

\AtBeginDocument{%
  }

\copyrightyear{2026}
\acmYear{2026}
\setcopyright{cc}
\setcctype{by-nc-nd}
\acmConference[CHI '26]{Proceedings of the 2026 CHI Conference on Human Factors in Computing Systems}{April 13--17, 2026}{Barcelona, Spain}
\acmBooktitle{Proceedings of the 2026 CHI Conference on Human Factors in Computing Systems (CHI '26), April 13--17, 2026, Barcelona, Spain}
\acmPrice{}
\acmDOI{10.1145/3772318.3790317}
\acmISBN{979-8-4007-2278-3/2026/04}
\begin{document}

\title{Fit Matters: Format--Distance Alignment Improves Conversational Search}

\author{Yitian Yang}
\orcid{0009-0000-7530-2116}
\affiliation{%
  \department{Computer Science}
  \institution{National University of Singapore}
  \city{Singapore}
  \country{Singapore}
}
\email{yang.yitian@u.nus.edu}

\author{Yugin Tan}
\orcid{0009-0006-7357-0436}
\affiliation{%
  \department{Computer Science}
  \institution{National University of Singapore}
  \city{Singapore}
  \country{Singapore}
}
\email{tanyugin@nus.edu.sg}

\author{Jung-Tai King}
\orcid{0000-0003-2745-5999}
\affiliation{%
  \department{Hua-Shih College of Education}
  \institution{National Dong Hwa University}
  \city{Hualien}
  \country{Taiwan}
}
\email{jtchin2@gms.ndhu.edu.tw}

\author{Yang Chen Lin}
\orcid{0000-0002-2477-4110}
\affiliation{%
  \department{Department of Computer Science}
  \institution{National Tsing Hua University}
  \city{Hsinchu}
  \country{Taiwan}
}
\email{yangchenlin@gapp.nthu.edu.tw}

\author{Yi-Chieh Lee}
\orcid{0000-0002-5484-6066}
\affiliation{%
  \department{Computer Science}
  \institution{National University of Singapore}
  \city{Singapore}
  \country{Singapore}
}
\email{yclee@nus.edu.sg}


\begin{abstract}
\input{Sections/abstract}
\end{abstract}




\maketitle
\input{Sections/introduction}
\input{Sections/related-work}
\input{Sections/current-study}
\input{Sections/method}
\input{Sections/result}
\input{Sections/discussion}
\input{Sections/conclusion}

\begin{acks}
This research is supported by the NUS Centre for Computational Social Science and Humanities (A-8002954), the NUS Artificial Intelligence Institute (A-8003879), and the special project of National Dong Hwa University (NDHU: 115T2951-10). We sincerely thank our participants for their participation and anonymous reviewers for their valuable comments and suggestions on this work.
\end{acks}

\bibliographystyle{ACM-Reference-Format}
\bibliography{reference}

\clearpage 

\end{document}

%% file: Sections/abstract.tex
Existing conversational search systems can synthesize information into responses, but they lack principled ways to adapt response formats to users' cognitive states. This paper investigates whether aligning format and distance, which involves matching information granularity and media to users' psychological distance, improves user experience. In a between-subjects experiment (N~=~464) on travel planning, we crossed two distance dimensions (temporal/spatial $\times$ near/far) with four formats varying in granularity (abstract/concrete) and media (text/image-and-text). The experiment established that format--distance alignment reduced users' risk perceptions while increasing decision confidence, perceptions of information usefulness, ease of use, enjoyment, and credibility, and adoption intentions. Concrete formats imposed higher cognitive load, but yielded productive effort when matched to near-distance tasks. Images enhanced concrete but not abstract text, suggesting multimedia benefits depend on complementarity. These findings establish format--distance alignment as a distinctive and important design dimension, enabling systems to tailor response formats to users' psychological distance.

%% file: Sections/introduction.tex
\section{Introduction}

Conversational search systems are fundamentally transforming online information seeking. Unlike traditional search engines that return ranked lists, artificial intelligence (AI)-powered systems like ChatGPT synthesize information into coherent conversational responses \cite{zamani2023conversational, radlinski2017theoretical}. This paradigm shift poses a critical design challenge: how to present information within synthesized responses. Consider a user asking about traveling to Paris. Should the system provide abstract overviews, or concrete details with specific hours and prices? Should it include images, or focus on text? While existing AI systems can generate varied formats \cite{achiam2023gpt, brown2020language}, they apply uniform formats in such cases, regardless of whether the user is exploring next year's vacation ideas or planning tomorrow's activities. In other words, they overlook how psychological distance -- people's subjective separation from their tasks -- systematically shapes information-processing preferences \cite{trope2010construal}.

None of this is to suggest that the designs of current conversational search systems are unsophisticated. Indeed, they are carefully crafted in multiple dimensions, including accuracy in information provision, anthropomorphism in interaction patterns, and personalization of content and style based on user profiles \cite{volkel2021eliciting, chaves2020should, zhang2018towards}. Nevertheless, they have rarely, if ever, operationalized \textit{format--distance alignment}: i.e., structurally adapted the modality and level of abstraction of the information to users' psychological distance. This oversight is particularly problematic in light of evidence that users' format preferences vary systematically with task context \cite{frummet2024decoding, hwang2023rewriting, vtyurina2017exploring}. Unlike personalization, which depends on user data and suffers from cold-start issues \cite{schein2002methods} and privacy concerns \cite{toch2012personalization}, format--distance alignment leverages task characteristics that can be systematically evaluated without any prior information about the user \cite{trope2010construal}. Research has shown that users exhibit divergent preferences for brief vs. detailed responses in conversational search \cite{sekulic2024towards, vtyurina2017exploring}, but most current systems treat these preference variations as random noise, rather than recognizing them as potentially predictable cognitive shifts.

Construal-level theory (CLT), a useful lens for understanding these preference patterns \cite{liberman2014traversing, trope2010construal}, posits that psychological distance fundamentally shapes how people process information. More specifically, tasks that evoke near distance (e.g., planning for tomorrow) trigger a preference for concrete details, while tasks that evoke far distance (e.g., planning for next year) trigger one for abstract overviews \cite{trope2003temporal, liberman1998role}. However, CLT has never been systematically applied to optimize conversational search responses, despite its success in explaining information processing across domains \cite{liberman2014traversing} and even recent evidence of its relevance to such searches' design. Notably, \citeauthor{yang2025understanding}~\cite{yang2025understanding} demonstrated that users' preference for conversational searches over traditional web searches was positively correlated with psychological distance. However, that study's focus was system-level preferences rather than format adaptation within conversational responses. A critical gap therefore remains: How should conversational search systems adjust their response formats based on users' psychological distance from their tasks?

This paper helps fill that gap through a controlled experiment (N~=~464) that examined how the alignment between information presentation formats (IPFs) and the subjects' psychological distance from their tasks affected users' experience of the responses and their cognitive processing during such search tasks. Using travel planning as a representative complex decision-making task \cite{choi2012structure, dellaert1998multi}, we manipulated psychological distance while systematically varying IPFs along two key dimensions: granularity (abstract vs. concrete) and media type (text vs. image-and-text).

The findings indicated that format--distance alignment consistently improved user experience across all measured dimensions. Surprisingly, concrete formats imposed a higher cognitive load but produced better outcomes when aligned with near-distance tasks \cite{bjork2011making}. Media-type effects manifested in a nuanced manner: concrete images substantially enhanced concrete information across all measured dimensions, whereas abstract image failed to yield similar benefits for abstract content. This challenges assumptions regarding the universal advantages of multimedia \cite{mayer2005cambridge}.

As such, this work makes three key contributions to HCI. First, it extends CLT-based research on modality choice between conversational and web search \cite{yang2025understanding} by showing that, beyond system selection, psychological distance systematically shapes format preferences within conversational search, and empirically validates format--distance alignment as a design dimension. Second, it shows that cognitive load and processing fluency operate independently, though higher processing demands can still coincide with superior user experience when information complexity matches users' cognitive orientation. And third, it offers actionable design principles for adaptive conversational searches: i.e., systems should detect psychological-distance cues to calibrate both their response granularity and their media-type selections.

%% file: Sections/related-work.tex
\section{Related Work}

\subsection{Construal-level Theory and Information Processing Fluency}
\label{sec:clt-ipf}

Information processing depends not only on objective content, but also on how individuals mentally construe it. CLT provides a theoretical framework for understanding this phenomenon \cite{trope2010construal}.

\textit{Psychological distance}, a core CLT concept, refers to a person's perception of the degree of separation between themselves and a target, which can be temporal (near vs. far future), spatial (nearby vs. distant locations), social (similar vs. dissimilar others), or hypothetical (likely vs. unlikely events) \cite{liberman2014traversing, trope2010construal}. These four dimensions share a common cognitive effect: as distance increases, mental representations systematically shift from context-specific details (low-level construals) toward generalized concepts (high-level construals) \cite{maglio2013distance, trope2010construal}; this construal shift fundamentally alters information processing. That is, even when facing an identical task objective, individuals construe it differently depending on psychological distance: those psychologically near focus on concrete features, practical constraints, and implementation details, whereas those psychologically distant privilege abstract principles, overall value, and overarching goals \cite{trope2010construal, trope2007construal}.

The cognitive implications of psychological distance shape people's \textit{processing fluency}, i.e., the subjective ease with which they process information cognitively \cite{reber2004processing}. When information granularity aligns with the receiving individual's construal level, their cognitive processing becomes more fluid \cite{schwarz2021metacognitive, lee2004bringing} and the effort required to understand different representational formats is reduced \cite{alter2009uniting}. Previous research has also revealed that the interaction between construal level and cognitive capacity strongly affects processing effectiveness \cite{korner2014concrete}. For example, when cognitive capacity is limited (e.g., due to time pressure), concrete construal facilitates processing through enhanced visualization and emotional engagement, making information more accessible and compelling. On the other hand, when cognitive capacity is ample, abstract construal enables more effective processing by directing attention to overarching principles and systematic relationships \cite{korner2014concrete, fujita2006construal}. This suggests that optimal information processing depends not on a fixed construal level, but on the alignment among psychological distance, available cognitive resources, and IPFs. Moreover, a ``good'' alignment not only reduces subjective difficulty, but also generates positive affect and increases decision confidence \cite{schwarz2004metacognitive, winkielman2003hedonic}.

Psychological distance is ubiquitous in daily life, holding ``fundamental and pervasive importance'' for cognition and behavior, and can therefore be assessed intuitively without deliberate effort \cite{trope2010construal}. This process is manifested in consumer behavior, where temporal distance systematically shifts product evaluations from feasibility to desirability features \cite{liberman2007construal}, and in online-communication contexts, where it shapes impression formation and messages' persuasiveness \cite{weidlich2024social, sordi2022construal}. These converging findings establish a robust empirical foundation for the idea that people's construal level fundamentally alters what types of information they find useful, convincing, and easy to process, and thus position CLT as able to inform testable design hypotheses about IPFs in conversational search.

\subsection{CLT-informed User-centered Design in Conversational Search Systems}

Conversational search systems are a critical application domain for examining user-centered design \cite{trippas2018informing, radlinski2017theoretical}, as they must meet diverse user needs and cognitive states across varying search contexts \cite{azzopardi2021cognitive, vakulenko2019qrfa}. Established approaches focus on factual accuracy \cite{zamani2023conversational, amershi2019guidelines}, anthropomorphism \cite{araujo2018living, verhagen2014virtual}, and personalization to individual preferences \cite{ha2024clochat, kocaballi2019personalization}. Personalization, which includes adaptations to content (e.g. topic-relevance algorithms) and style (e.g. tone, register, and linguistic patterns) \cite{chaves2020should, li2016persona}, has proven beneficial for user satisfaction and trust. However, such personalization approaches face inherent limitations, particularly privacy concerns \cite{toch2012personalization} and cold-start problems \cite{schein2002methods} arising from their reliance on user data and profiling. Format--distance alignment, while still overlooked, avoids these issues and may in fact serve as a useful complement to existing personalization methods.

Published evidence on conversational searches reveals clear patterns in user preferences that prevailing design frameworks cannot explain. \citeauthor{frummet2024decoding}~\cite{frummet2024decoding}, for instance, found that in procedural conversational tasks such as cooking, users generally prefer sentence-length conversational responses, but that in time-sensitive steps (e.g., heating in oil), they shift toward very brief, directive answers. At a broader level, \citeauthor{vtyurina2017exploring}~\cite{vtyurina2017exploring} observed fluctuations in user preferences for brief, direct answers vs. more detailed explanations, which existing approaches treat as random variation. 
\citeauthor{yang2025understanding}~\cite{yang2025understanding} suggested that this was because abstract, synthesized responses better matched people's construal needs at greater psychological distances. However, while AI-powered conversational search systems are technically capable of generating responses with varying information granularity and multiple media types~\cite{koh2023generating, wu2023multimodal}, existing systems do not systematically adapt these format dimensions based on users' psychological distance. This study aims to address this gap by proposing format--distance alignment as a design principle to manage these variations and by investigating the impacts of such adaptive systems.

\subsection{Information Presentation Formats in Conversational Search Systems}

Creating a meaningful theoretical framework for format--distance alignment requires a systematic examination of how conversational search systems' various presentation formats map onto people's construal levels. To do this, we identify two fundamental dimensions of IPFs: information granularity \cite{trope2010construal} and media type \cite{amit2009distance, shneiderman2003eyes}.

\subsubsection{Information Granularity: Abstract vs. Concrete}

Information granularity determines content's specificity and level of detail. At the abstract end of the granularity spectrum, information is synthesized into key points and overarching concepts, providing high-level conceptual views that facilitate pattern recognition and rapid relevance assessment \cite{marchionini2006exploratory, hearst1995tilebars}. This abstraction proves particularly valuable during exploratory searches, when users need to navigate complex information spaces without being overwhelmed by details \cite{jonassen2013structural, chi2002framework}. While abstraction aids comprehension efficiency, however, it may omit critical decision factors, compromising people's evaluative capacity \cite{payne1993adaptive, keller1987effects}.

Conversely, concrete formats deliver fine-grained details, including numerical data, specific examples, and procedural instructions. This specificity improves not only the precision but also the actionability of the information presented \cite{brandstatter2001implementation}, and can thus be seen as essential for time-sensitive situations in which reduced uncertainty directly impacts decision confidence \cite{wu2022time, crescenzi2021adaptation}. However, while concrete information supports immediate task execution, its rich detail can induce information overload, particularly among users who have not yet reached clear evaluative criteria \cite{eppler2004concept}.

\subsubsection{Media Type: Text vs. Image and Text}

Differing media types trigger distinct cognitive-processing pathways, with prior research suggesting that text tends to promote abstract, category-based processing and images tend to evoke concrete, perceptual details \cite{amit2012you, amit2009distance}. However, this dichotomy is far from cut and dried, as both text and images can convey information at varying abstraction levels. Text can range from high-level thematic overviews and conceptual frameworks to detailed specifications and step-by-step procedures \cite{trope2010construal}, while images run the gamut from schematic representations and simple symbols to richly detailed illustrations and photographs of complex subject matter \cite{butcher2006learning, schnotz2003construction}. This flexibility represents a key challenge for attempts at format--distance alignment, due to the difficulty of quickly or intuitively matching people's construal levels to multimodal presentations' abstraction levels. The situation is further complicated by the critical dependence of different media types' effectiveness on their alignment with processing goals and task demands \cite{schnotz2003construction, mayer2002multimedia}.

Ideally, multimodal presentations leverage the complementary strengths of text, which provides logical structure and precise factual information, and images, which offer visual context, spatial understanding, and emotional resonance \cite{schnotz2003construction}. However, empirical evidence indicates that multimedia benefits are highly conditional. Health-communication researchers, for example, have demonstrated that images improve comprehension only when directly relevant to instructions, and that decorative visuals actually impair readers' understanding \cite{houts2006role}. This conditional effectiveness extends to data visualization, where graphical representations excel at revealing trends, but often fail to convey precise values \cite{larkin1987diagram, cleveland1984graphical}.

\subsubsection{Integrating Format Dimensions}

The interaction between information granularity and media type creates complex design challenges for conversational systems. As discussed in Section ~\ref{sec:clt-ipf}, processing fluency emerges from alignment between information format and users' construal level \cite{alter2009uniting}, so when multiple format dimensions vary simultaneously, achieving this alignment becomes increasingly complex.

The ramifications of the fact that both textual and visual modalities vary in granularity are likewise important. That is, when abstraction levels are well-matched between text and visual elements (e.g. abstract diagrams are paired with conceptual summaries, or concrete photographs with detailed instructions), their formats reinforce each other and thereby enhance cognitive processing \cite{schnotz2003construction}. However, incongruent combinations (such as abstract text paired with concrete images) create representational conflicts that increase cognitive load and impair comprehension \cite{butcher2006learning, schnotz2003construction}. Given that misaligned abstraction levels inherently impair user experience, we isolate the effects of multimedia by focusing exclusively on congruent combinations where text and visual abstraction levels match.

All of this raises fundamental questions about format effects in conversational searches. First, to establish the baseline effects of media type, we ask:

\textbf{RQ1.} How do different combinations of media type (text vs. image-and-text) and information granularity (abstract vs. concrete) influence user outcomes in conversational searches, particularly when images and text maintain congruent levels of abstraction?

Building on the answers to RQ1, we can proceed to an examination of format--distance alignment, which -- following CLT principles -- we define as the correspondence between IPF characteristics and users' construal levels. We deem \textit{matched conditions} to occur when format granularity aligns with psychological distance: concrete formats for proximal tasks and abstract formats for distal tasks. \textit{Mismatched conditions}, meanwhile, present the opposite pattern. If format--distance alignment enhances processing fluency, as theoretically predicted, we should observe consistent improvements in user experience metrics under matched conditions. Thus:

\textbf{RQ2.} How does the match or mismatch between IPFs and users' psychological distance from their tasks affect user outcomes?

%% file: Sections/current-study.tex
\section{Current Study and Hypotheses}

\subsection{Travel Planning as a Representative Task Scenario}

To validate the effects of matching IPFs with psychological distance, we selected a travel-planning task of sufficient complexity to be considered relevant in the real world. Travel planning has been widely adopted as a testbed in HCI research, including work on how psychological distance affects information preferences \cite{yang2025understanding} and interface requirements \cite{zhang2016designing}. More importantly, travel decisions naturally involve multiple psychological-distance dimensions, such as \textit{when} (temporal distance) and \textit{to where} (spatial distance) a trip is taking place. Although CLT encompasses four distance dimensions, the psychological-commonality principle suggests that these dimensions produce consistent cognitive effects \cite{maglio2013distance, trope2010construal}. While recent work has validated cross-dimensional consistency \cite{yang2025understanding}, it remains crucial to test whether specific format adaptations provide consistent benefits across these dimensions. We therefore manipulated both temporal and spatial distances to examine potential interaction effects and to verify the stability of the format--distance alignment across distance types.

\subsection{Cognitive Processing and User Experience}

Although the effects of media-type combinations are an unsettled topic of our current exploration (RQ1), CLT, cognitive-load theory, and processing fluency provide a clear foundation for developing hypotheses about format--distance alignment (RQ2). Based on these three theoretical frameworks, we propose eight hypotheses organized around two complementary mechanisms: cognitive-resource allocation (H1-H2) and user perceptions (H3-H8).

\subsubsection{Cognitive Load and Decision Confidence in Information-seeking Tasks}
\label{sec:cl-dc}

In general, processing concrete information inherently requires more cognitive resources due to its detailed nature \cite{kirschner2002cognitive}. We expect this pattern to hold across all IPFs in conversational searches. Therefore:

\textbf{H1. In conversational search results, concrete information will result in a higher cognitive load than abstract information.}

However, cognitive load in information processing is not uniformly detrimental. Following the distinction between germane and extraneous load \cite{sweller2011cognitive}, we recognize that mental effort may be productive when aligned with task demands. To distinguish the nature of these load states, we draw on \textit{Cognitive Closure Theory} \cite{kruglanski1996motivated}. Productive cognitive load fosters \textit{cognitive flexibility}, encouraging users to explore and weigh alternatives. Conversely, load stemming from format misalignment motivates users to terminate mental effort quickly. This manifests as a ``seizing and freezing'' strategy, leading to \textit{premature cognitive closure}. We therefore additionally analyze these distinct mechanisms through specific linguistic markers in user outputs.

People's confidence in their decisions reflects a combination of certainty about, and satisfaction with, their choices \cite{moore2008trouble, sniezek1992groups}. In plain language, ``IPFs matching construal levels'' means that information is being provided in the form best suited to users' current cognitive needs. More specifically, near-distance tasks are well-matched with concrete, detail-rich formats that address feasibility and implementation, whereas far-distance tasks are well-matched with abstract overviews that support high-level planning and value trade-offs \cite{trope2010construal, trope2007construal}. Such alignment improves processing fluency and reduces perceived uncertainty, which in turn raises decision confidence \cite{alter2009uniting, schwarz2004metacognitive}. Therefore:

\textbf{H2. When IPF's information granularity matches psychological distance, users will exhibit higher decision confidence.}

\subsubsection{Information Adoption Model in Information Presentation Formats}
\label{sec:iam}

While H1 and H2 cover the cognitive mechanisms of format--distance alignment, a rounded understanding of user acceptance will also require us to examine how processing fluency translates into subjective evaluations and behavioral intentions. We therefore adopt the information adoption model (IAM) \cite{sussman2003informational} as this study's main framework for capturing those experiential dimensions, and couple it with an additional factor, enjoyment \cite{winkielman2003hedonic, davis1992extrinsic}, to account for the hedonic value of fluent information processing.

IAM and its extensions have been widely used to explain how users evaluate and adopt information provided by systems \cite{erkan2016influence, bhattacherjee2006influence}. Our variant of IAM builds directly on the processing-fluency mechanism described in section ~\ref{sec:clt-ipf}. When information abstraction aligns with users' psychological distance, processing becomes more fluid \cite{reber2004processing}, and these fluency gains are unique for each dimension of user perception. In addition, the elimination of translation overhead fundamentally reduces the subjective effort users experience during information processing \cite{gefen2000relative}. Accordingly:

\textbf{H3. When IPF's information granularity matches psychological distance, perceived ease of use will increase.}

As explained above, matched formats deliver optimally relevant information for task contexts \cite{sussman2003informational, gefen2000relative}. Thus:

\textbf{H4. When IPF's information granularity matches psychological distance, perceived usefulness will increase.}

Processing fluency extends its influence beyond utilitarian dimensions by generating positive affect through metacognitive ease \cite{reber2004processing, winkielman2003hedonic}. This hedonic response is particularly crucial to the success of conversational interfaces, where sustained engagement depends on users' intrinsic satisfaction with the interaction experience \cite{venkatesh2000theoretical, davis1992extrinsic, van2004user}. Therefore:

\textbf{H5. When IPF's information granularity matches psychological distance, enjoyment will increase.}

Format mismatches tend to amplify perceived risk \cite{tseng2016perceived} by reducing processing fluency and increasing subjective uncertainty \cite{schwarz2004metacognitive, alter2009uniting}. Thus:

\textbf{H6. When IPF's information granularity matches psychological distance, perceived risk will decrease.}

One of the functions of format--distance alignment is to signal systems' contextualized awareness of users' actual informational needs. This apparent competence enhances source trustworthiness and information believability \cite{fogg2001makes, acikgoz2024integrated, rabjohn2008examining}. Accordingly:

\textbf{H7. When IPF's information granularity matches psychological distance, perceived information credibility will increase.}

Lastly, IAM predicts that improvements in perceived ease of use, usefulness, and information credibility should converge with lower perceived risk to collectively strengthen users' information-adoption intentions \cite{erkan2016influence, cheung2008impact}. Thus:

\textbf{H8. When IPF's information granularity matches psychological distance, users' intention to use the information will increase.}

%% file: Sections/method.tex
\section{Methodology}

This study employed a between-subjects experiment to examine how IPFs and psychological distance influence users' experience with the responses of conversational searches. We manipulated psychological distance (temporal vs. spatial; near vs. far) and IPFs (varying in media type and granularity) in a 2 $\times$ 2 $\times$ 4 factorial design. The following subsections describe our experimental system, design, participants, procedure, measurements, and analytical approach.

\begin{figure*}[htbp!]
  \centering
  \includegraphics[width= 0.9\textwidth]{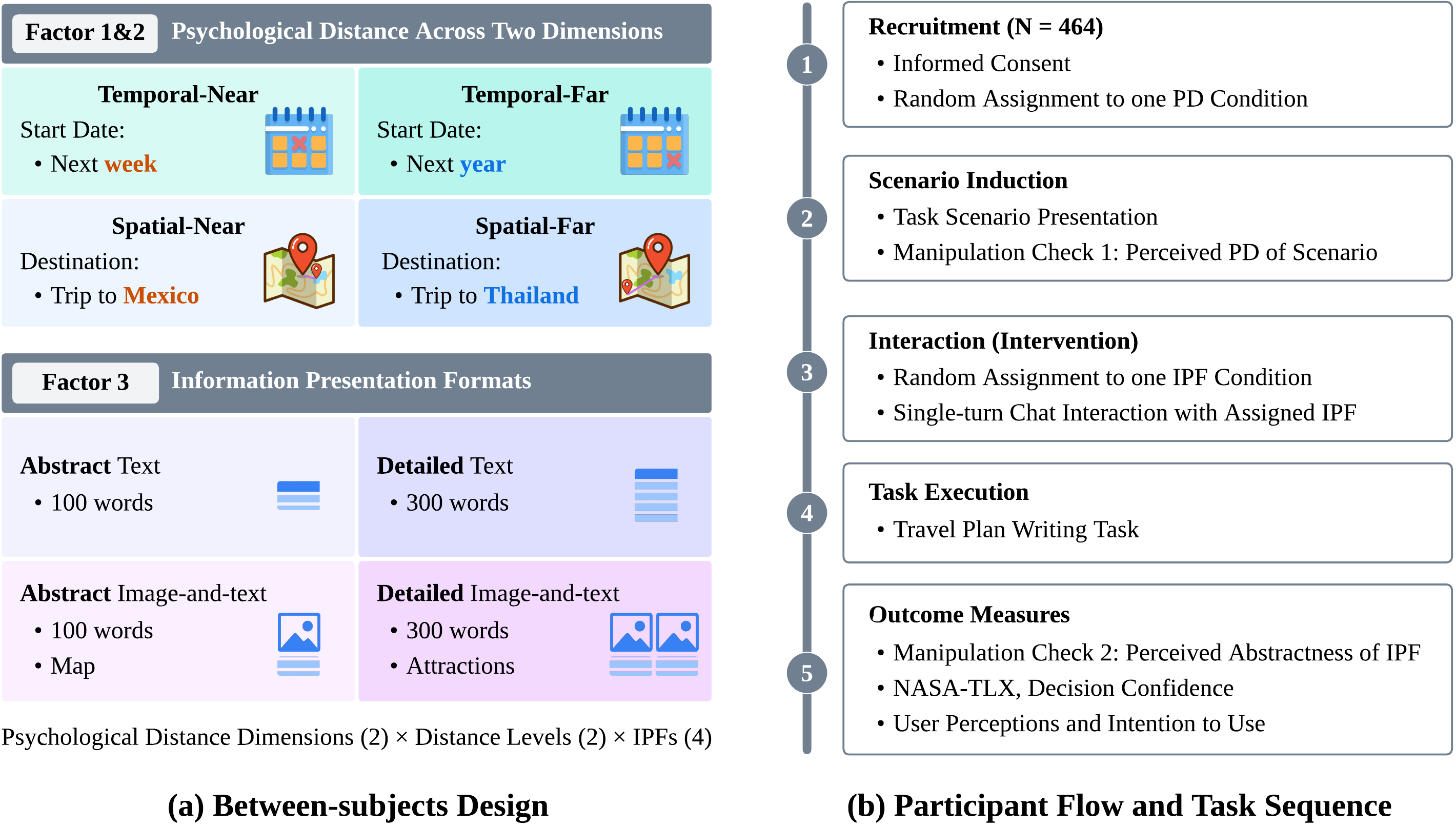}
  \Description{Schematic overview of the experimental design and participant flow. The left panel shows a 2 by 2 by 4 between-subjects design. Participants are first randomly assigned to one of four psychological-distance conditions, combining temporal versus spatial distance with near versus far scenarios. Temporal conditions involve trips to Mexico either next week or next year. Spatial conditions involve trips next month either to Mexico or to Thailand. Within each psychological-distance condition, participants are randomly assigned to one of four information presentation formats: abstract text of about one hundred words, detailed text of about three hundred words, abstract image-and-text that combines an abstract summary with an overview map, and detailed image-and-text that combines a detailed description with three photographs of specific attractions. The right panel shows the sequence of study steps that all participants follow. After consent, participants view the scenario, complete manipulation checks and background travel questions, interact once with the simulated chat interface that presents the assigned format, write a short travel plan, and then complete post-task measures including an abstraction check, NASA-TLX cognitive load, decision confidence, user perceptions of the information format, and intention to use the format.}
  \caption{Overview of the experimental design and participant flow. The left panel (a) summarizes the 2 $\times$ 2 $\times$ 4 between-subjects structure that crosses psychological distance (temporal vs.\ spatial $\times$ near vs.\ far) with four information presentation formats (abstract text, detailed text, abstract image-and-text, detailed image-and-text), yielding sixteen experimental conditions. The right panel (b) depicts the common sequence of steps that each participant completes, from consent and scenario presentation through manipulation checks, the single-turn chat interaction with the assigned format, the travel-plan writing task, and the post-task outcome measures. 
  \emph{Note.} PD = psychological distance; IPF = information presentation format.}
  \label{fig:design-flow}
\end{figure*}

\subsection{System Design}

We developed a controlled experimental system that replicated ChatGPT's conversational interface through a custom HTML/CSS framework embedded in Qualtrics \footnote{\url{https://www.qualtrics.com/support/survey-platform/getting-started/survey-platform-overview/}}. Users initiated queries through a familiar chat-input field, and responses were delivered using a streaming text animation to simulate the natural cadence of conversational AI responses \cite{wang2024task}. These characteristics supported ecological validity by replicating the temporal dynamics of actual ChatGPT interactions, while also ensuring that all participants received information at identical rates, to guard against the confounding factors that might arise from variation in their reading speeds \cite{rayner1998eye}. There was then a 10-second waiting period to further ensure that all participants had fully read the information before being allowed to proceed to the next step.

\subsection{Experimental Design}

As summarized in Figure~\ref{fig:design-flow}(a), participants were randomly assigned first to one of four psychological-distance conditions (temporal/spatial $\times$ near/far), then to one of four IPFs (abstract text, detailed text, abstract image-and-text, detailed image-and-text), resulting in 16 participant groups, each under a unique experimental condition.\footnote{Complete experimental materials, including all scenario scripts and stimulus examples, are provided in Appendix C.}

\subsubsection{Psychological Distance Manipulations}
In line with CLT \cite{trope2010construal}, we operationalized psychological distance through task scenarios that manipulated either temporal or spatial dimensions while holding other factors constant. We elected to employ imagined manipulations, based on meta-analytic evidence of their stronger effects compared to ostensibly real manipulations \cite{soderberg2015effects}.

\textbf{Temporal Distance Manipulation.} Participants in the conditions that experienced temporal-distance manipulations were presented with scenarios about planning a trip to Mexico with varying time horizons. The temporal-near condition stated: ``You are an avid global traveler who takes trips almost every month. You have been planning a vacation to Mexico that could take place either next week or next year. After considering your options, you have decided to go \textit{next week}.'' The temporal-far condition specified ``\textit{next year}'' with otherwise identical framing. All temporal conditions received Mexico-based IPFs.

\textbf{Spatial Distance Manipulation.} We operationalized spatial distance using participants' U.S. residency as a reference point. Neighboring Mexico represented spatial proximity, while trans-Pacific Thailand represented spatial distance. Participants in the spatial-distance manipulation conditions received scenarios with a fixed temporal frame, ``next month'', as follows: ``You are an avid global traveler who takes trips almost every month. For your next month's vacation, you considered both Mexico and Thailand as potential destinations. After weighing your options, you have decided on [\textit{Mexico/Thailand}] for this trip.'' Spatial-near participants received Mexico-based content; spatial-far participants received Thailand-based content. We developed parallel information sets for both destinations with equivalent content across cultural, culinary, and natural-attraction categories.

\subsubsection{Information Presentation Formats}

The four IPFs we created hereafter will be referred to as abstract text (AT), detailed text (DT), abstract image-and-text (AIAT), and detailed image-and-text (DIDT). All textual content was generated using GPT-4o \cite{hwang2023rewriting} with carefully designed prompts.

For the two abstract formats, we generated summaries of approximately 100 words emphasizing high-level themes. AT highlighted cultural significance, culinary diversity, and natural features without operational details. AIAT paired this text with a regional overview map that marked major tourist areas rather than specific attractions, offering a coarse spatial summary of the destination. This presentation aligns with CLT’s characterization of abstract construals as high-level, structural representations \cite{alter2009uniting}.

For the two concrete formats, we produced approximately 300-word responses containing precise operational information. DT included exact opening hours, admission prices, specific addresses, and quantitative specifications (e.g., ``El Castillo pyramid stands 24 meters tall with 91 steps on each of four sides''). DIDT supplemented this text with photographs of specific attractions, providing rich perceptual information that complemented the textual specifications.

All eight format-destination combinations (4 IPFs $\times$ 2 destinations) underwent piloting to ensure that textual descriptions maintained comparable clarity across conditions and that visual stimuli accurately instantiated the intended construal levels \footnote{Appendix C.2.2 shows reduced-size thumbnails of these images; in the study interface, all maps and photographs were rendered at a size that allowed participants to read labels and inspect key details.}. This systematic approach ensured clean manipulation of our theoretical dimensions while maintaining external validity through genuine AI-generated travel information.

\subsection{Participants}

Participants were recruited through CloudResearch Connect\footnote{\url{https://www.cloudresearch.com/products/connect-for-researchers/}}, with residency restricted to the United States to ensure the uniformity of their spatial-distance perceptions of other countries. After removing 32 participants who failed attention checks, the final sample consisted of 464. The 16 experimental groups each had 27 to 30 members ($M = 29.0$, $SD = 0.97$). 

There were 234 participants who identified as female (50.4\%), 226 who identified as male (48.7\%), and four who had another gender identification or preferred not to disclose it (0.9\%). Participants' ages ranged from 18 upward ($M_{age} = 33.8$, $SD = 9.2$). They were well-educated, with 73.5\% holding at least a bachelor's degree. The study lasted approximately 20 minutes, and each participant who passed the attention checks received compensation of \$3.50. The study protocol was approved by the National University of Singapore Departmental Ethics Review Committee.

\subsection{Procedure}

Participants completed the study entirely online through Qualtrics. As shown in Figure~\ref{fig:design-flow}(b), after providing informed consent, they were randomly assigned to one of four psychological-distance conditions and presented with their travel-planning scenario. 

Then, as a manipulation check, they rated perceived psychological distance on a seven-point scale. To control for potential confounds, the participants then reported two aspects of their travel experience: their typical travel frequency and their familiarity with the assigned destination (Mexico/Thailand).

The participants then engaged with the core experimental task through our simulated ChatGPT interface. A pre-configured prompt appeared automatically (``Hi, I'm going to [destination] [timeframe]. What should I know before I go?''), which participants initiated by clicking ``Send''. The system then delivered its travel information in one of the four formats described above, assigned at random, and no further conversational turns were allowed so that participants’ perceptions could be attributed solely to the presented response format. After receiving it, participants were asked to write a travel plan of at least 25 words, which both ensured active engagement with the stimuli and approximated a common pattern where users consolidate online search results into personal notes or itineraries.

The post-task assessment phase collected all primary outcome measures: 1) a manipulation check on perceived information abstractness, using a seven-point scale; 2) NASA Task Load Index (NASA-TLX) measurement of cognitive load across six dimensions; 3) three items on decision confidence; and 4) rating of the IPF for usefulness, ease of use, enjoyment, information credibility, perceived risk, and intention to use. 

\subsection{Measurements}

We employed validated multi-item scales to assess cognitive load, decision confidence, and user perceptions of IPFs. Additionally, we analyzed participants' written travel plans to derive objective linguistic indicators of their cognitive processing styles. \footnote{Complete measurement instruments, including all item wordings and response scales, are provided in Appendix B.
Internal consistency reliabilities (Cronbach's $\alpha$) for all scales are reported in Appendix A.2.}

\subsubsection{Cognitive Load}
Our instrument for assessing cognitive load, NASA-TLX \cite{hart1988development}, is a multidimensional scale capturing mental effort during information processing. In its raw form, it comprises six subscales rated from 0 to 100, i.e., mental demand, physical demand, temporal demand, performance (reverse-scored, where 0~=~Perfect, 100~=~Failure), effort, and frustration. Following recommendations for experimental contexts \cite{hart2006nasa}, we computed an unweighted composite score averaging all six dimensions.

\subsubsection{Decision Confidence}
Our measurement of the participants' certainty regarding their travel-planning choices comprised three items that were responded to on 7-point Likert scales ranging from 1~=~strongly disagree to 7~=~strongly agree \cite{o1995validation}. There were ``I feel confident that I made the best choice based on this information,'' ``I have no doubts about the quality of my decision,'' and ``I am certain that my decision is correct.''

\subsubsection{User Perceptions}

Building on the IAM \cite{sussman2003informational} and technology acceptance model \cite{davis1989perceived}, we assessed the following five perceptual dimensions of IPFs.

\begin{itemize}
    \item \textbf{Perceived Ease of Use}. We captured the cognitive effort required to process the presented information using four items  \cite{davis1989perceived}: ``I find the information easy to understand'', ``It is easy to obtain the information I need from this format'', ``Interaction with this format is clear and understandable'', and ``Learning to use this format would be easy''.
    \item \textbf{Perceived Usefulness}. The extent to which the participants believed the conversational-search format enhanced their travel-planning effectiveness was measured via an additional four items \cite{davis1989perceived}: ``Using this format would enhance my travel-planning effectiveness'', ``This format improves my efficiency when planning a trip'', ``I find this information format useful'', and ``Using this format would increase my productivity in researching travel''.
    \item \textbf{Enjoyment}. To capture the search experience's hedonic value, as an extension of traditional information-acceptance measures, we used three items \cite{davis1992extrinsic}: ``Using this format is fun'', ``I enjoy using this information format'', and ``I find interacting with this format pleasant''.
    \item \textbf{Perceived Information Credibility}. To assess the participants' perceptions that the conversational-search system was trustworthy and accurate, we adopted four items from information-credibility research \cite{appelman2016measuring}: ``The information provided is accurate'', ``The information is reliable'', ``I can trust the information'', and ``Overall, the information is credible''.
    \item \textbf{Perceived Risk}. Finally, to capture the participants' concerns about relying on the presented format as the basis of travel decisions, we adopted four items \cite{featherman2003predicting}: ``It would be risky to rely on this information format for travel planning'', ``Using this format could lead to negative consequences for my trip'', ``I feel uncertain about the outcomes of using this information format'', and ``This information format is safe to use''.
\end{itemize}

\subsubsection{Intention to Use}

A single item measured the participants' willingness to adopt the presented search format for their future travel planning. Following IAM conventions \cite{cheung2008impact, sussman2003informational}, it was ``I would prefer information presented in this format for my travel planning'', rated on a five-point scale ranging from 1~=~strongly disagree to 5~=~strongly agree.

\subsubsection{Travel Plan Analysis}

To complement self-reported metrics with objective behavioral data, we analyzed participants' written travel plans using a computational linguistic approach. Informed by the Linguistic Inquiry and Word Count (LIWC) framework \cite{pennebaker2015development}, we developed two domain-specific lexicons to capture distinct cognitive strategies under uncertainty:
\begin{itemize}
    \item \textbf{Exploratory Markers}: Words indicating openness to alternatives and active deliberation (e.g., ``might,'' ``could,'' ``perhaps,'' ``maybe,'' ``possibly''). Higher frequency suggests users are engaging in \textit{cognitive flexibility} \cite{kruglanski1996motivated}, effectively weighing options provided by the system.
    \item \textbf{Closure Markers}: Words indicating fixed decisions and a desire for finality (e.g., ``will,'' ``must,'' ``definitely,'' ``certainly,'' ``booking,'' ``booked''). In the context of complex information processing, excessive use of these markers can signal \textit{premature cognitive closure} -- a strategy to terminate cognitively taxing tasks by seizing on early solutions \cite{kruglanski1996motivated, webster1994individual}.
\end{itemize}

We calculated the frequency of these markers in each valid travel plan to operationalize cognitive engagement styles.

\subsubsection{Manipulation Checks}

We verified experimental manipulations through two items. Psychological distance was measured via the participant's rating of ``Overall, how near or far did the travel task feel to you?'' on a seven-point scale from 1~=~very near to 7~=~very far. Information abstraction was assessed post-task using ``Thinking about the TRAVEL information (search result) you just received, how would you describe it?'', rated on a seven-point scale from 1~=~very abstract to 7~=~very concrete.

\subsection{Analytical Approach}

We employed a hierarchical analytical strategy following established conventions for factorial designs. All analyses used two-tailed tests ($\alpha = .05$) with Benjamini-Hochberg FDR (BH-FDR) correction within hypothesis families \cite{benjamini1995controlling}. For all ANOVAs with significant coefficients, we conducted simple effects analyses using independent-samples $t$-tests, applying Welch's correction when Levene's test indicated unequal variances ($p < .05$). Effect sizes were reported as $\eta_p^2$ for ANOVAs and as Cohen's $d$ for pairwise comparisons.

\subsubsection{Manipulation Validation}

The robustness of psychological-distance manipulations was validated using independent-samples $t$-tests (near vs. far), conducted separately for the temporal and spatial dimensions. Information-abstraction levels across the four IPFs were verified through one-way ANOVAs with Tukey's HSD post hoc comparisons.

\subsubsection{Hypothesis Testing}

\begin{itemize}
    \item \textbf{Cognitive Load (H1)}. We first examined the main effect of information granularity (abstract vs. concrete) on composite NASA-TLX scores using one-way ANOVA. To assess the complete factorial structure, we also tested for distance main effects and granularity $\times$ distance interactions. Simple effects analyses were then used to compare granularity effects within each distance condition (near and far separately).
    \item \textbf{Format--Distance Alignment (RQ2, H2–H8).} We coded alignment as a binary variable comprising matched conditions (abstract--far, concrete--near) vs. mismatched ones (abstract--near, concrete--far). The alignment effect was tested through the distance $\times$ granularity interaction term in 2$\times$2 factorial ANOVAs. Stratified analyses within granularity levels were then applied to ascertain whether alignment effects operated symmetrically across abstraction levels. We applied an analogous two-factor structure to the behavioral indicators derived from the travel plans and regarded these analyses as an exploratory triangulation of the self-reported outcomes.
\end{itemize}

\subsubsection{Media Effects (RQ1)}

The main effects of text-only vs. image-and-text presentations were tested using one-way ANOVAs across all user-experience dimensions, while media type $\times$ granularity interactions were examined through 2 $\times$ 2 factorial ANOVAs. Significant interactions were decomposed through simple effects analyses that compared media types within each granularity level.

\subsubsection{Robustness Analyses}
We conducted two robustness checks to address potential confounds in the spatial distance manipulation, namely how participants' individual experiences with traveling (to any country in general) may have affected their receptiveness to and influence by AI-generated information\footnote{Detailed robustness check results are reported in Appendix A.3.}.
First, we employed ANCOVA models that included travel frequency and prior destination visits as covariates. The IPFs $\times$ Distance interaction remained significant for all user-experience outcomes after BH–FDR correction ($p_{BH}$ $<$ .01). In contrast, cognitive load showed no interaction effect ($p_{BH}$ = .42), a finding consistent with H1 that cognitive load operates independently of format alignment. Second, we examined the spatial-distance subsample specifically by incorporating destination (Mexico vs. Thailand) as an additional between-subjects factor. Format--distance alignment benefits persisted across both destinations. These results confirm that our effects were not driven by idiosyncratic characteristics of either travel target.

%% file: Sections/result.tex
\section{Results}

Our manipulation checks confirmed that our experimental conditions successfully induced differential psychological-distance and information-granularity perceptions. Our subsequent analyses then revealed robust evidence for format--distance alignment effects (RQ2). Specifically, aligning information granularity with users' psychological distance systematically improved decision confidence (H2), perceived ease of use (H3), usefulness (H4), enjoyment (H5), information credibility (H7), and intention to use (H8), while reducing perceived risk (H6). Finally, addressing RQ1, we demonstrated that media types differentially shaped user experience: when concrete images were paired with concrete text, they increased decision confidence, perceived ease of use, enjoyment, and intention to use, while simultaneously reducing perceived risk; by contrast, abstract images paired with abstract text yielded no comparable gains. Unless otherwise noted, $p$ values reported below are BH–FDR adjusted within the corresponding outcome family.

\subsection{Manipulation Checks}

\noindent \textbf{Psychological Distance.} Our manipulation-check analyses confirmed that participants in the \textit{far} distance conditions perceived their scenarios as significantly more distant than those in the \textit{near} conditions, in both the temporal and spatial dimensions. On the temporal side, participants who planned a trip \textit{next year} (M = 4.73, SD = 1.22) rated the event as significantly farther away in time than those who planned a trip \textit{next week} (M = 3.21, SD = 1.95), $t(191.10) = 7.09$, $p < .001$***, Cohen's $d = 0.93$. Similarly, for spatial distance tasks, participants planning a trip to \textit{Thailand} ($M = 4.73$, $SD = 2.18$) perceived significantly greater geographical distance than those planning a trip to \textit{Mexico} ($M = 3.82$, $SD = 1.63$), $t(213.08) = 3.62$, $p < .001$***, Cohen's $d = 0.48$. 
These results confirm that psychological distance manipulations were effective.

Preliminary analyses for moderation revealed that no interactions between the type of distance dimension, psychological distance, information granularity, or media type significantly influenced any outcome variable (all $p$s $> .150$).\footnote{Complete results for the omnibus three-way and four-way interaction effects are reported in Appendix A.1.}
This absence of moderation effects justified our subsequent analyses' collapsing of temporal and spatial dimensions into a unified theoretical construct, psychological distance, consistent with CLT.

\noindent \textbf{Information Granularity.} We next demonstrate through a one-way ANOVA that, as intended, half the IPFs differed in their abstraction level from the other half ($F(3, 460) = 313.33$, $p < .001$***). Post hoc Tukey comparisons revealed that participants who received DT ($M = 6.06$, $SD = 1.24$) rated the information as significantly more concrete than those who received either AT ($M = 2.12$, $SD = 1.27$), mean difference $= 3.94$, $p < .001$***, or AIAT ($M = 2.48$, $SD = 1.46$), mean difference $= 3.58$, $p < .001$ ***. Similarly, participants who received DIDT ($M = 6.04$, $SD = 1.31$) perceived the information as significantly more concrete than those who received AT, mean difference $= 3.92$, $p < .001$***, or AIAT, mean difference $= 3.57$, $p < .001$***. No significant differences were found between the two detailed formats (DT vs. DIDT), $p = 1.00$, or between the two abstract formats (AT vs. AIAT), $p = .172$. These results confirm that our manipulation of information granularity was successful, with detailed formats perceived as significantly more concrete than abstract formats.

\begin{table*}[htbp]
\centering
\caption{Statistical Results for the Matching Effect Between Information Granularity and Psychological Distance}
\label{tab:matching-effects}
\begin{tabular}{lccccccc}
\toprule
Dependent Variable & $F(1, 460)$ & $p$ & $\eta_p^2$ & Matched $M(SD)$ & Mismatched $M(SD)$ & Cohen's $d$ & Outcome\\
\midrule
Decision Confidence (H2) & 18.03 & $< .001$*** & .038 & 5.42 (1.27) & 4.88 (1.42) & 0.40 & Support \\
Ease of Use (H3) & 23.61 & $< .001$*** & .049 & 5.59 (1.15) & 5.01 (1.39) & 0.45 & Support \\
Usefulness (H4) & 26.39 & $< .001$*** & .054 & 4.99 (1.40) & 4.32 (1.57) & 0.45 & Support \\
Enjoyment (H5) & 46.18 & $< .001$*** & .091 & 4.89 (1.34) & 3.97 (1.53) & 0.63 & Support \\
Risk\textsuperscript{a} (H6) & 13.67 & $< .001$*** & .029 & 3.19 (1.26) & 3.63 (1.28) & 0.34 & Support \\
Information Credibility (H7) & 14.60 & $< .001$*** & .031 & 5.56 (0.83) & 5.25 (0.90) & 0.36 & Support \\
Intention to Use (H8) & 24.78 & $< .001$*** & .051 & 3.39 (1.23) & 2.81 (1.33) & 0.45 & Support \\
\bottomrule
\end{tabular}
\vspace{2pt}
\caption*{\normalfont\footnotesize
\parbox{0.92\linewidth}{
\raggedright
\textit{Note}. $F$ values represent the granularity $\times$ distance interaction term from 2 $\times$ 2 factorial analyses of variance. The matched conditions are abstract--far and concrete--near; and the mismatched conditions ones, abstract--near and concrete--far. *$p < .05$, **$p < .01$, ***$p < .001$.\\
\textsuperscript{a} In this variable, lower values indicate better outcomes.
}}
\end{table*}

\subsection{Concrete Information Increased Cognitive Load (H1)}

We hypothesized (H1) that concrete information would result in higher cognitive load than abstract information did. A one-way ANOVA examining the effect of information granularity on cognitive load revealed a significant main effect, $F(1, 462) = 13.10$, $p < .001$***, $\eta_p^2 = .028$, directly supporting H1. Participants processing concrete information ($M = 38.69$, $SD = 18.19$) experienced a significantly higher cognitive load than those processing its abstract counterpart ($M = 32.70$, $SD = 17.49$). 
In other words, as compared to abstract travel information, the detailed versions demanded substantially more mental effort, working memory, and attentional resources.

\begin{figure}[htbp!]
    \centering
    \includegraphics[width=\linewidth]{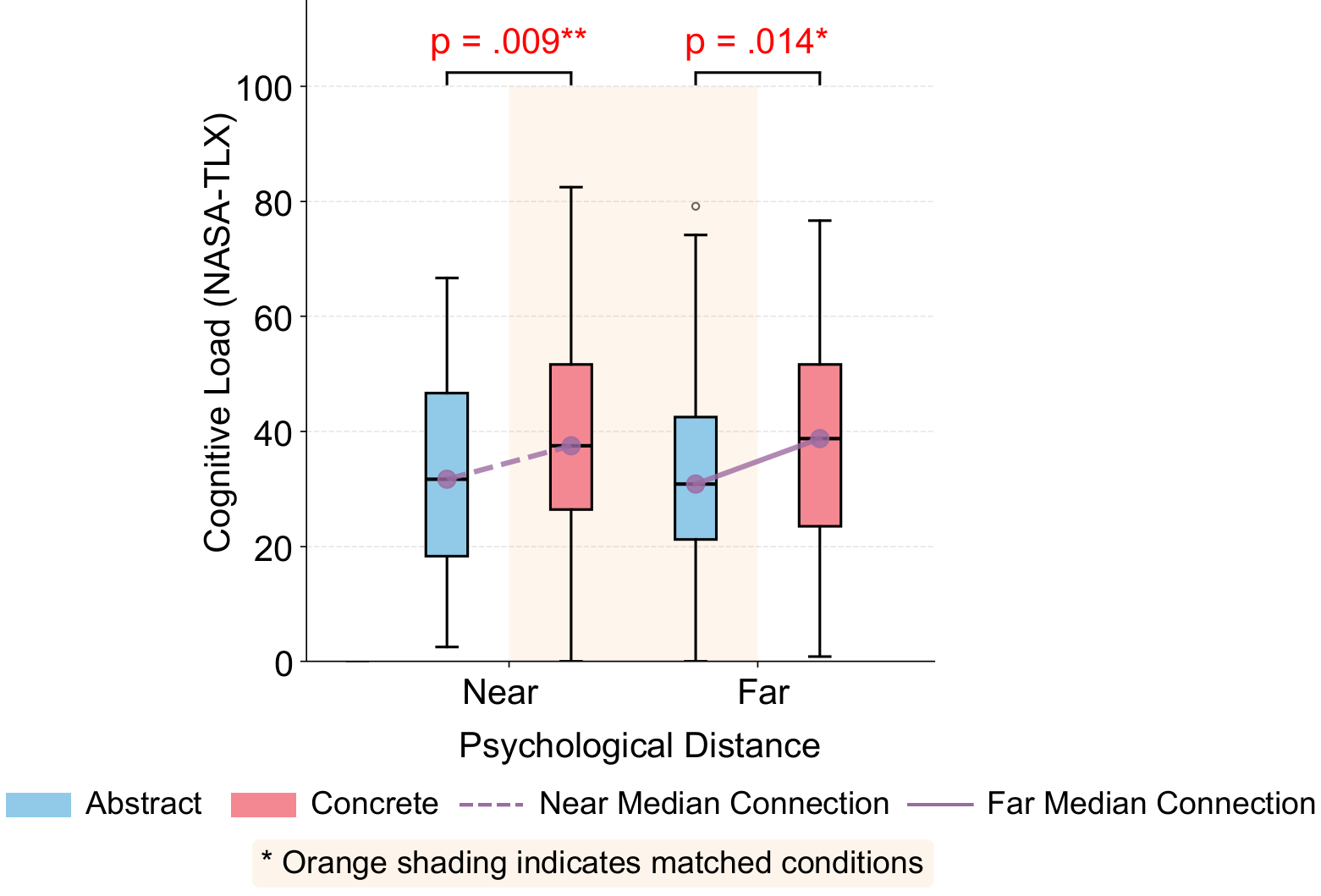}
    \Description{Box plots of NASA–TLX cognitive load by psychological distance (Near, Far) and information granularity (Abstract, Concrete). The y-axis is cognitive load from 0–100; the x-axis groups Near and Far, each with two box plots (Abstract and Concrete). Orange shading marks theoretically matched cells (Abstract--Far, Concrete--Near). Within Near, Concrete shows a higher median load than Abstract (p = .009); within Far, Concrete is also higher than Abstract (p = .014). Overall, matched cells have lower medians than the mismatched cells (Abstract--Near, Concrete--Far). Medians are linked by guide lines; box plots show median, interquartile range, whiskers to range, and any outliers.}
    \caption{Cognitive load (NASA-TLX) as a function of psychological distance and information granularity. Lower scores indicate reduced cognitive burden. Box plots show median, quartiles, and data range. The lines connecting the Near/Far group medians are provided as visual aids and were not used in statistical analyses. Significance levels: *p < .05, **p < .01.}
    \label{fig:cognitive-load-nasa}
\end{figure}

As Figure ~\ref{fig:cognitive-load-nasa} illustrates, this pattern was similar across both psychological-distance conditions. For near-distance scenarios, concrete information ($M = 39.08$, $SD = 17.99$) resulted in significantly higher cognitive load than abstract information did ($M = 32.86$, $SD = 17.84$), $t(229) = 2.64$, $p = .009$**, $d = 0.35$. Similarly, for far-distance scenarios, concrete information ($M = 38.31$, $SD = 18.44$) produced a higher cognitive load than abstract information ($M = 32.54$, $SD = 17.22$), $t(230.58) = 2.47$, $p = .014$*, $d = 0.32$.

However, neither the main effect of psychological distance ($p = .757$) nor the interaction between distance and information granularity ($p = .893$) was statistically significant. These results suggest that information granularity, rather than psychological distance or (mis)matching conditions, was the primary driver of cognitive load. Therefore, while matching may benefit other user-experience outcomes, it should not be expected to alleviate the inherent cognitive demands of processing detailed information.

\subsection{The Effect of Matching Information Granularity with Psychological Distance (RQ2, H2–H8)}
\label{result:effectOfMatching}

Aligning information granularity with users' psychological distance significantly improved their conversational-search outcomes. As Table ~\ref{tab:matching-effects} shows, when abstract information matched far distance, or concrete information matched near distance, users experienced enhanced decision confidence, perceived ease of use, perceived usefulness, enjoyment, perceived information credibility, and intention to use, while also perceiving risk as lower ($p < .001$*** across all outcomes). These consistent outcomes demonstrate that cognitive fit between information abstraction and psychological construal fundamentally shapes user experience of conversational searches.

To examine these effects in greater detail, we conducted follow-up analyses using 1) independent-samples $t$-tests, to compare matched vs. mismatched conditions for each outcome variable; and 2) paired comparisons within each information-granularity level to understand the directional patterns of the interactions. The results revealed two distinct patterns: symmetric benefits of matching that operated bidirectionally across most user-experience dimensions, and asymmetric sensitivities that manifested differently for decision confidence and risk perception. Each is explained in its own subsection below.

\subsubsection{Benefits of Matching that Operated Symmetrically across User Experience Dimensions}

\begin{figure*}[htbp!]
   \centering
   \includegraphics[width=0.95\textwidth]{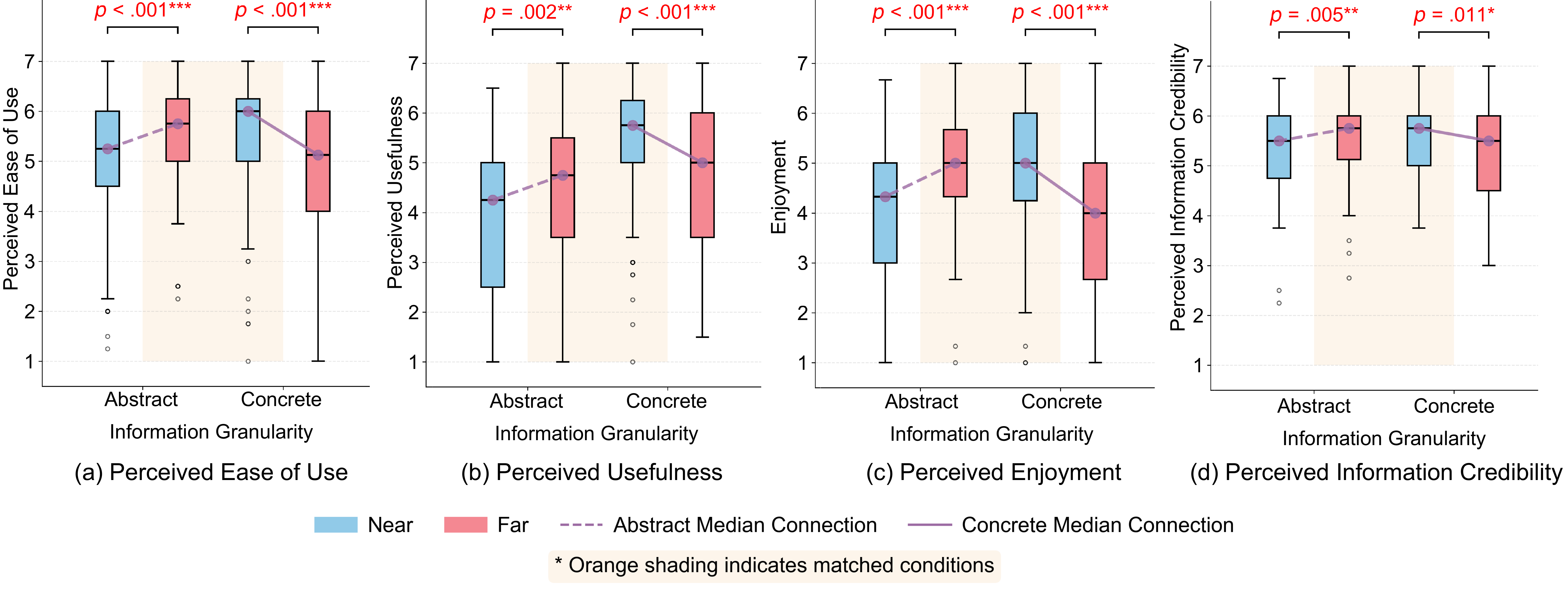}
   \Description{Four boxplot panels show user experience ratings by information granularity and psychological distance. Each panel plots 1–7 ratings on the y-axis against Abstract and Concrete on the x-axis, with Near in blue and Far in red. Orange shading marks the theoretically matched cells Abstract with Far and Concrete with Near. Medians are linked by guide lines, dashed for Abstract and solid for Concrete. Panel (a) Ease of Use shows matched conditions rated higher, Abstract p<.001 and Concrete p<.001. Panel (b) Usefulness shows matched higher, Abstract p=.002 and Concrete p<.001. Panel (c) Enjoyment shows matched higher, Abstract p<.001 and Concrete p<.001. Panel (d) Information Credibility shows matched higher, Abstract p=.005 and Concrete p=.011. Box plots display medians, interquartile ranges, whiskers to non-outlier ranges, and individual outliers.}
   \caption{Symmetric matching effects of information granularity and psychological distance alignment. Matched conditions consistently outperformed mismatched ones across all four user-experience dimensions. Box plots show median, quartiles, and data range. The lines connecting the Near/Far group medians are provided as visual aids and were not used in statistical analyses. Significance levels: *p < .05, **p < .01, ***p < .001.}
   \label{fig:symmetric-effects}
\end{figure*}

As we have seen, aligning information granularity with psychological distance produced consistent improvements across multiple user-experience dimensions. Table ~\ref{tab:ux-matching-effects} and Figure ~\ref{fig:symmetric-effects} show that, as well as being symmetrical, these benefits manifested bidirectionally. That is, abstract information enhanced user experience for far-distance tasks while concrete information improved the outcomes for near-distance ones.

\begin{table*}[htbp!]
\centering
\caption{Effects of Matching across All User Experience Dimensions for Abstract and Concrete Information}
\label{tab:ux-matching-effects}
\begin{tabular}{llcccccc}
\toprule
Info Granularity & Dependent Variable & Matched $M(SD)$ & Mismatched $M(SD)$ & $t$ & $df$ & $p$ & Cohen's $d$ \\
\midrule
\multirow{5}{*}{Abstract\textsuperscript{a}}
& Ease of Use & 5.62 (0.99) & 5.09 (1.33) & 3.42 & 211.49 & $<$ .001*** & 0.45 \\
& Usefulness & 4.48 (1.41) & 3.88 (1.54) & 3.11 & 228 & .002** & 0.41 \\
& Enjoyment & 4.88 (1.15) & 4.04 (1.41) & 4.91 & 219.34 & $<$ .001*** & 0.65 \\
& Information Credibility & 5.56 (0.87) & 5.24 (0.86) & 2.84 & 228 & .005** & 0.37 \\
& Intention to Use & 3.11 (1.24) & 2.54 (1.15) & 3.64 & 228 & $<$ .001*** & 0.48 \\
\midrule
\multirow{5}{*}{Concrete\textsuperscript{b}}
& Ease of Use & 5.55 (1.28) & 4.93 (1.45) & 3.50 & 229.40 & $<$ .001*** & 0.46 \\
& Usefulness & 5.49 (1.19) & 4.74 (1.49) & 4.26 & 222.85 & $<$ .001*** & 0.56 \\
& Enjoyment & 4.90 (1.51) & 3.91 (1.64) & 4.80 & 232 & $<$ .001*** & 0.63 \\
& Information Credibility & 5.56 (0.80) & 5.26 (0.95) & 2.57 & 226.66 & .011* & 0.34 \\
& Intention to Use & 3.66 (1.16) & 3.08 (1.44) & 3.45 & 223.23 & $<$ .001*** & 0.45 \\
\bottomrule
\end{tabular}
\vspace{2pt}
\caption*{\normalfont\footnotesize
\parbox{0.92\linewidth}{
\raggedright
\textit{Note}. Degrees of freedom reflect Student's $t$ (equal variances) or Welch's $t$ (unequal variances). *$p < .05$, **$p < .01$, ***$p < .001$.\\
\textsuperscript{a,b} ``Matched'' and ``mismatched'' are defined relative to psychological distance. For category (a), matched = far and mismatched = near; for category (b), matched = near and mismatched = far.
}}
\end{table*}

The symmetry of these effects provides compelling evidence in favor of their underlying mechanism being processing fluency. Enjoyment exhibited the strongest matching benefits ($d$ = .63-.65), suggesting that cognitive alignment generates positive affect beyond information value. Similarly, the robust effects that emerged for perceived usefulness ($d$ = .41-.56) and ease of use ($d$ = .45-.46), indicated that format--distance alignment fundamentally improved both the subjective experience of information processing and the perceived utility of the content itself.

The consistency we observed across dimensions, ranging from the utilitarian (e.g., usefulness) to the affective (enjoyment), indicates that matching operates through a domain-general mechanism rather than isolated cognitive or emotional pathways. Even information credibility, which has traditionally been viewed as mostly content-dependent, was significantly higher under matched conditions ($d$ = .34-.37), suggesting that appropriate abstraction levels enhance trust. Finally, participants reported significantly higher intended-use scores in matched than in mismatched conditions ($d$ = .45-.48). This improvement from perceptual improvements to adoption intentions should be sufficient to establish format--distance alignment as a fundamental principle for conversational-search design rather than a peripheral optimization.

\subsubsection{Asymmetric Patterns: in Decision Confidence and Risk Perception}

\begin{figure}[htbp!]
   \centering
   \includegraphics[width=\linewidth]{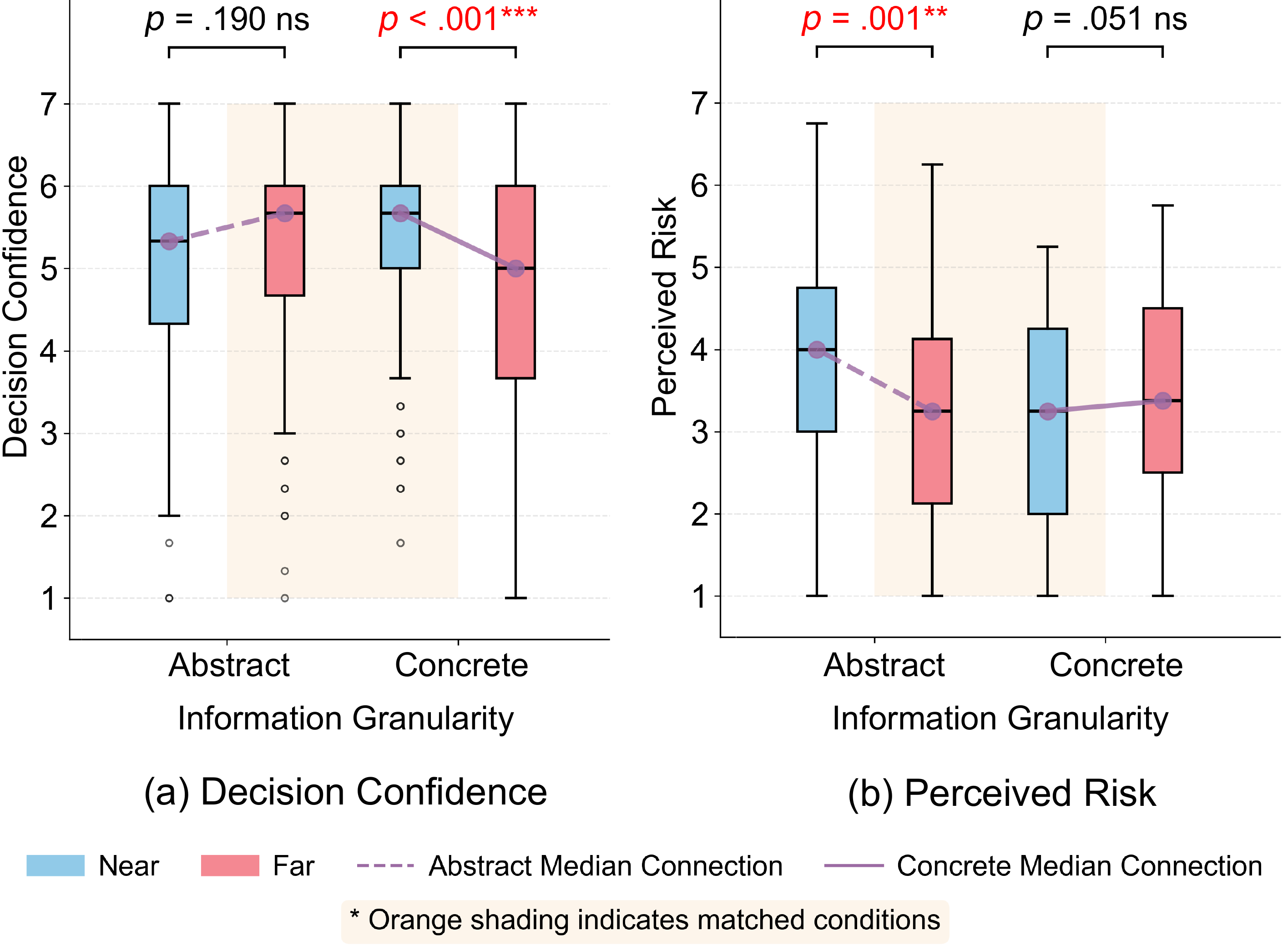}
   \Description{Two boxplot panels show asymmetric matching effects for decision confidence and perceived risk by information granularity and psychological distance. Panel (a) plots decision confidence ratings from 1 to 7 on the y-axis against Abstract and Concrete information on the x-axis, with Near in blue and Far in red. In Abstract, there is no significant difference between Near and Far with p=.190. In Concrete, Near shows higher confidence than Far with p<.001. Panel (b) plots perceived risk ratings from 1 to 7 on the y-axis against Abstract and Concrete information on the x-axis. In Abstract, Near shows higher perceived risk than Far with p=.001. In Concrete, the difference is not significant with p=.051. Orange shading highlights the theoretically matched cells Abstract with Far and Concrete with Near. Medians are connected by guide lines, dashed for Abstract and solid for Concrete. Box plots display medians, interquartile ranges, whiskers to non-outlier ranges, and individual outliers.}
   \caption{Asymmetric matching effects revealing differential sensitivity to format--distance misalignment. Panel (a) shows that decision confidence benefits exclusively from concrete--near matching, and is unaffected by abstract information. Panel (b) shows that risk perception increases only when abstract information mismatches with near distance, and that concrete information maintains stable risk across distances. These unidirectional patterns contrast with the bidirectional benefits observed in the other dimensions. Box plots show median, quartiles, and data range. The lines connecting the Near/Far group medians are provided as visual aids and were not used in statistical analyses. Significance levels: **p < .01, ***p < .001; ns = not significant.}
   \label{fig:asymmetric-effects}
\end{figure}

While most user experience dimensions exhibited symmetric matching benefits, decision confidence and risk perception revealed distinct asymmetric patterns that suggest differential cognitive mechanisms governing evaluative judgments under uncertainty. These asymmetries, visualized in Figure ~\ref{fig:asymmetric-effects} and quantified in Table ~\ref{tab:asymmetric-effects}, demonstrate that information granularity selectively influences confidence formation and risk assessment through distinct psychological pathways.

\begin{table*}[htbp]
\centering
\caption{Asymmetric Effects of Matching When Information is Abstract vs. Concrete}
\label{tab:asymmetric-effects}
\begin{tabular}{llcccccc}
\toprule
Info Granularity & Dependent Variable & Matched $M(SD)$ & Mismatched $M(SD)$ & $t$ & $df$ & $p$ & Cohen's $d$ \\
\midrule
\multirow{2}{*}{Abstract\textsuperscript{a}} 
& Decision Confidence & 5.32 (1.38) & 5.08 (1.40) & 1.32 & 228 & .190 & 0.17 \\
& Risk\textsuperscript{c} & 3.25 (1.29) & 3.80 (1.31) & 3.23 & 228 & .001** & 0.43 \\
\midrule
\multirow{2}{*}{Concrete\textsuperscript{b}} 
& Decision Confidence & 5.51 (1.13) & 4.69 (1.41) & 4.88 & 222.90 & $<$ .001*** & 0.64 \\
& Risk\textsuperscript{c} & 3.14 (1.23) & 3.46 (1.24) & 1.96 & 232 & .051 & 0.26 \\
\bottomrule
\end{tabular}
\vspace{2pt}
\caption*{\normalfont\footnotesize
\parbox{0.9\linewidth}{
\raggedright
\textit{Note}. Degrees of freedom reflect Student's $t$ (equal variances) or Welch's $t$ (unequal variances). *$p < .05$, **$p < .01$, ***$p < .001$.\\
\textsuperscript{a,b} ``Matched'' and ``mismatched'' are defined relative to psychological distance. For category (a), matched = far and mismatched = near; for category (b), matched = near and mismatched = far. \\
\textsuperscript{c} In this variable, lower values indicate better outcomes.
}}
\end{table*}

Decision confidence showed pronounced sensitivity to concrete information alignment ($d$ = 0.64), but the alignment of abstract information produced negligible effects on it. This pattern suggests that when psychological distance is near, concrete specifications provide the operational certainty necessary for confident decision-making. The absence of matching benefits for abstract information, meanwhile, indicates that high-level overviews retain people's baseline confidence regardless of psychological distance.

Risk perception exhibited the complementary asymmetry, responding selectively to abstract information misalignment. That is, when abstract overviews were presented for near-distance tasks, the participants recognized the operational gaps that threatened successful execution, resulting in significantly elevated risk perceptions ($d$ = .43). Conversely, concrete information was associated with relatively stable risk perceptions across psychological-distance conditions ($d$ = .26), suggesting that excessive detail represents inefficiency rather than threat when psychological distance is far.

These asymmetric patterns reveal how specific information types selectively engage distinct evaluative mechanisms, depending on users' psychological distance from their tasks.

\subsubsection{Format--Distance Alignment Influences Cognitive Processing Styles}

Analysis of participants' written travel plans revealed significant interaction effects that provide behavioral evidence for the cognitive mechanisms underlying format-distance alignment benefits.

This analysis provided behavioral corroboration for the observed format--distance alignment effects. A significant Distance $\times$ Granularity interaction emerged for the use of \textit{exploratory markers}, $F(1, 460) = 4.20$, $p = .041$*, $\eta_p^2 = .009$. Participants in matched conditions employed significantly more exploratory language ($M = 0.22$, $SD = 0.48$) compared to those in mismatched conditions ($M = 0.12$, $SD = 0.35$), $t(462) = 2.51$, $p = .012$, $d = 0.24$. This indicates that when information format aligned with psychological distance, participants were more likely to use language associated with weighing options and considering alternatives.

Conversely, a significant interaction was found for \textit{closure markers}, $F(1, 460) = 4.24$, $p = .040$*, $\eta_p^2 = .009$. Participants in mismatched conditions used significantly more definitive closure language ($M = 1.33$, $SD = 1.21$) than those in matched conditions ($M = 1.06$, $SD = 1.08$), $t(462) = 2.54$, $p = .011$, $d = 0.24$. Notably, this increased use of definitive language in mismatched conditions occurred despite these participants reporting lower decision confidence (see Section \ref{result:effectOfMatching}), revealing a divergence between linguistic expression and subjective certainty.

\begin{figure}[htbp!]
  \centering
  \includegraphics[width=0.8\linewidth]{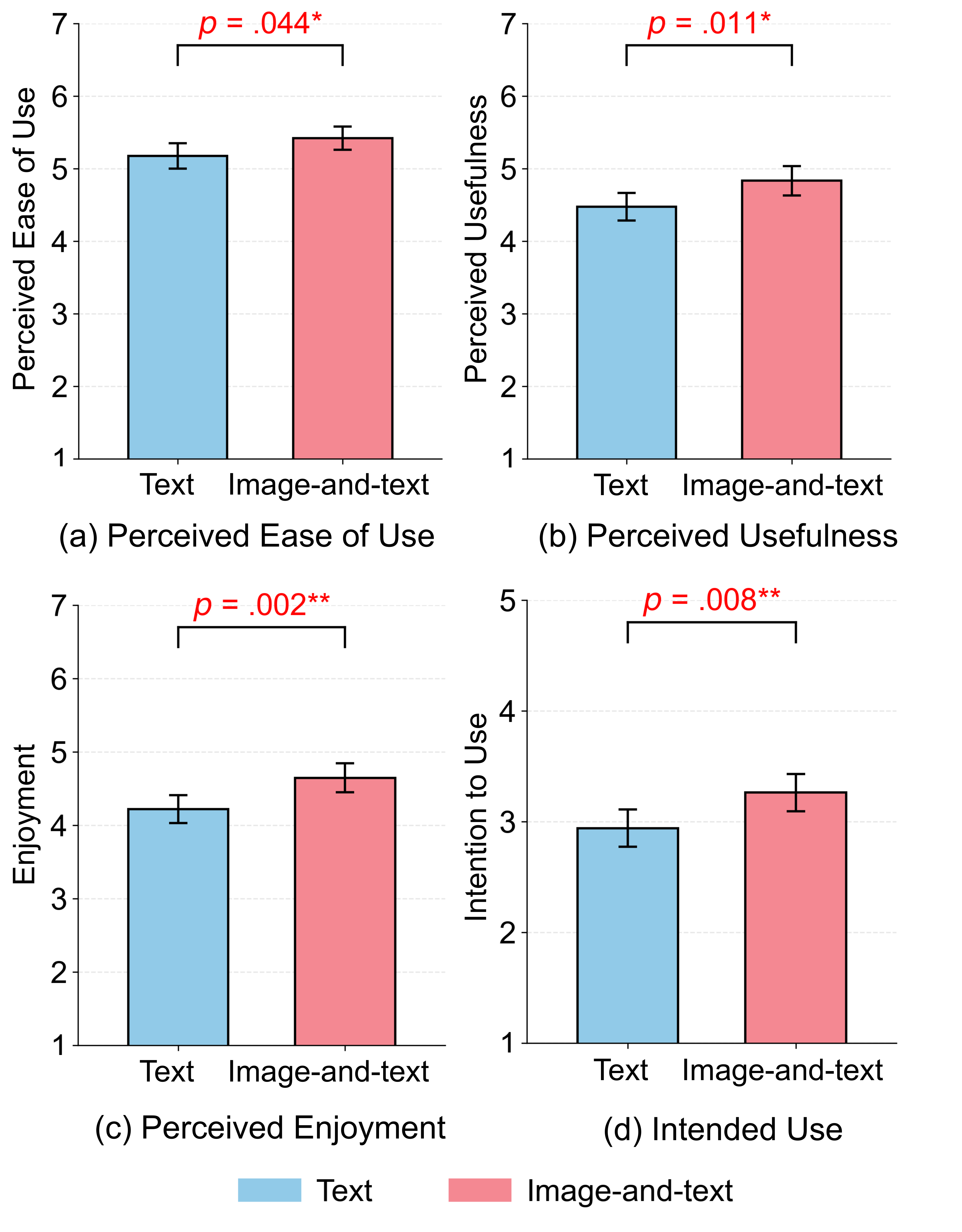}
  \Description{Four bar-chart panels compare media formats on four user-experience outcomes. Each panel plots ratings from 1–7 on the y-axis with two bars on the x-axis: Text in blue and Image-and-text in red. Error bars indicate 95\% confidence intervals. Results show a consistent advantage for Image-and-text. Panel (a) Perceived Ease of Use: Image-and-text exceeds Text with p=.044. Panel (b) Perceived Usefulness: Image-and-text exceeds Text with p=.011. Panel (c) Enjoyment: Image-and-text exceeds Text with p=.002. Panel (d) Intention to Use: Image-and-text exceeds Text with p=.008.}
  \caption{Main effects of media type on user-experience outcomes. Error bars represent 95\% confidence intervals. Significance levels: *p < .05, **p < .01.}
  \label{fig:boxpanel-mediatype}
\end{figure}

\subsection{Visual Elements Enhanced Concrete but Not Abstract Information (RQ1)}

Image-and-text formats outperformed text-only presentations for perceived ease of use, usefulness, enjoyment, and intention to use. However, this effect was significantly impacted by information granularity. That is, the inclusion of visual elements substantially enhanced concrete information across all four above-mentioned dimensions, while also reducing perceived risk. However, it provided minimal benefits when content was abstract, implying that multimodal presentation's effectiveness depends critically on the presented information's level of detail.

\subsubsection{Image-and-Text Formats Enhanced User Experience across Multiple Dimensions}

Figure ~\ref{fig:boxpanel-mediatype} visualizes the results of our examination of the main effects of text vs. image-and-text on user-experience outcomes. Its panels (a)-(d) show the four outcomes for which the advantages of image-and-text formats consistently outperformed text-only formats across four key user-experience dimensions.

Multimodal presentation was rated higher than text-only presentation for perceived ease of use ($M = 5.42$, $SD = 1.24$ vs. $M = 5.17$, $SD = 1.36$), $F(1, 462) = 4.09$, $p = .044$*, $\eta_p^2 = .009$, suggesting that \textbf{visual elements facilitate information processing} by providing cognitive scaffolding. This advantage also extended to perceived usefulness (image-and-text: $M = 4.83$, $SD = 1.55$ vs. text-only: $M = 4.47$, $SD = 1.48$), $F(1, 462) = 6.49$, $p = .011$*, $\eta_p^2 = .014$, indicating that visual enrichment enhances task-support effectiveness. Multimodal formats also generated substantially higher enjoyment than text-only formats (image-and-text: $M = 4.65$, $SD = 1.52$ vs. text-only: $M = 4.22$, $SD = 1.48$), $F(1, 462) = 9.51$, $p = .002$**, $\eta_p^2 = .020$; and this translated into behavioral intentions, as participants reported higher intention to use image-and-text formats than text-only ones ($M = 3.26$, $SD = 1.30$ vs. $M = 2.94$, $SD = 1.31$), $F(1, 462) = 7.03$, $p = .008$**, $\eta_p^2 = .015$.

Notably, however, media type had no significant effects on cognitive load ($p = .260$), decision confidence ($p = .338$), perceived risk ($p = .469$), or perceived information credibility ($p = .995$). These non-significant results suggest that, while images enhance the user experience along engagement and utility dimensions, they neither reduce the mental effort required to process information nor influence users' trust or certainty about their decisions.

\subsubsection{Image-and-Text Formats Amplified the Benefits of Concrete but Not Abstract Information}

\begin{table}[htbp]
  \centering
  \caption{Interaction effects of media type and information granularity on user-experience outcomes.}
  \label{tab:media-type-granularity-interactions}
  \begin{tabular}{lccc}
    \toprule
    Dependent Variable & $F(1, 460)$ & $p$ & $\eta_p^2$ \\
    \midrule
    Ease of Use & 21.22 & $p < .001$*** & .044 \\
    Enjoyment & 3.92 & $p = .048$* & .008 \\
    Risk\textsuperscript{a} & 6.55 & $p = .011$* & .014 \\
    Intention to Use & 4.60 & $p = .032$* & .010 \\
    Decision Confidence & 5.91 & $p = .015$* & .013 \\
    \bottomrule
  \end{tabular}
  \vspace{2pt}
  \caption*{\normalfont\footnotesize
  \parbox{0.82\linewidth}{
  \raggedright
  \textit{Note}. All tests are media-type by information-granularity interactions. *$p < .05$, **$p < .01$, ***$p < .001$\\
  \textsuperscript{a} In this variable, Lower values indicate better outcomes.
  }}
\end{table}

\begin{figure*}[htbp!]
   \centering
   \includegraphics[width=0.8\textwidth]{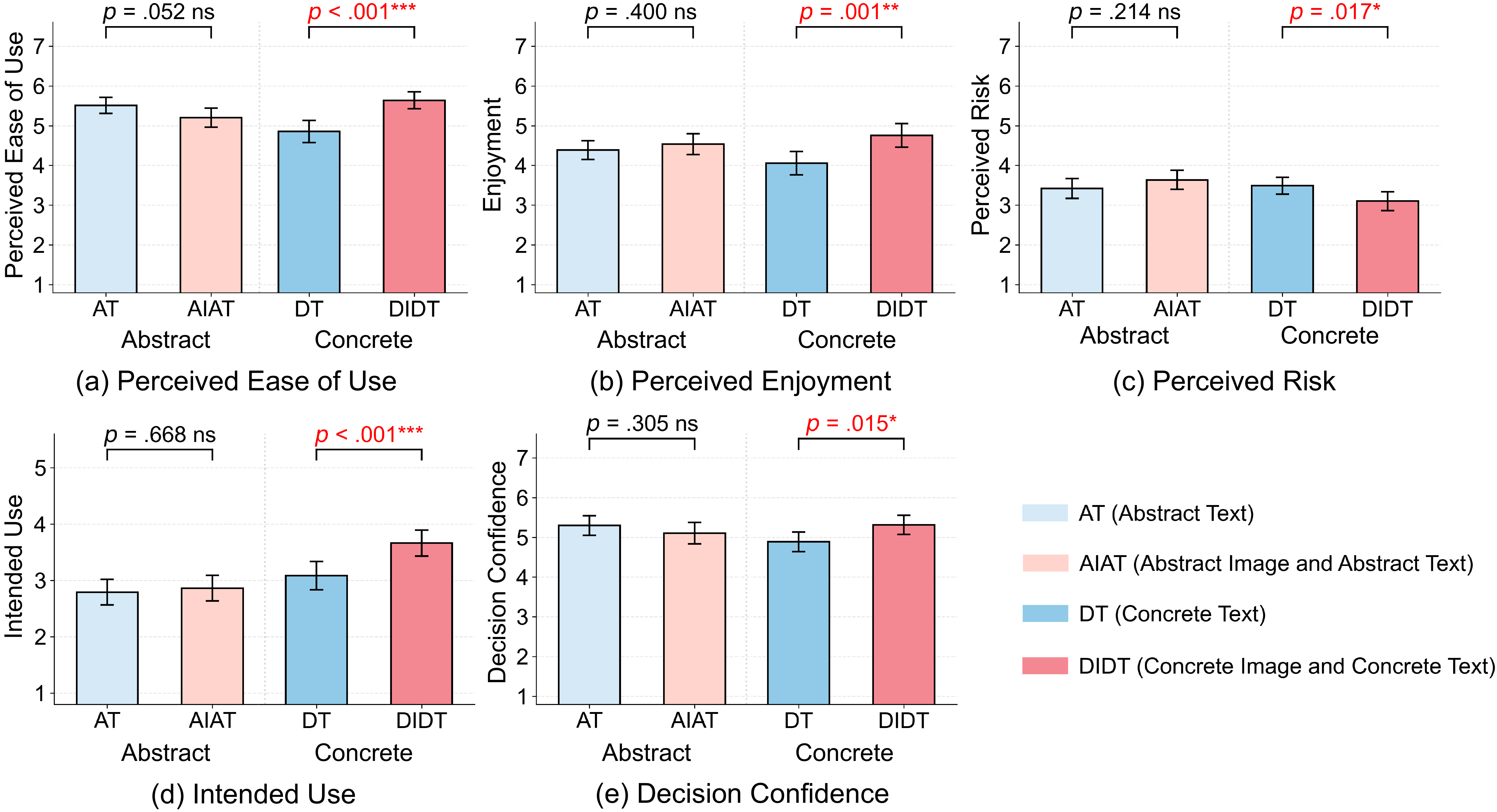}
   \Description{Five panel bar charts show media type by information granularity effects on five outcomes. Each panel plots ratings from 1 to 7 on the y-axis with four bars on the x-axis: AT (abstract text, light blue), AIAT (abstract image and text, light red), DT (concrete text, blue), and DIDT (concrete image and text, red). Error bars indicate 95\% confidence intervals. Panel (a) Perceived Ease of Use shows no difference between AT and AIAT with p=.051, and DIDT higher than DT with p<.001. Panel (b) Perceived Enjoyment shows no abstract difference with p=.400, and DIDT higher than DT with p=.001. Panel (c) Perceived Risk shows no abstract effect with p=.214, and DIDT lower than DT with p=.016. Panel (d) Intention to Use shows no abstract effect with p=.668, and DIDT higher than DT with p<.001. Panel (e) Decision Confidence shows no abstract effect with p=.305, and DIDT higher than DT with p=.015. Across panels, abstract comparisons are not significant while concrete comparisons favor image and text.}
   \caption{Interaction effects between media type and information granularity, revealing differential benefits of visual elements. This asymmetric pattern demonstrates that visual elements enhance user experience specifically when paired with detailed content. Error bars represent 95\% confidence intervals. Significance levels: *p < .05, **p < .01, ***p < .001; ns = not significant.}
   \label{fig:interaction-mediatype-granularity}
\end{figure*}

Media type and information granularity produced significant interactions across multiple user-experience outcomes (Table~\ref{tab:media-type-granularity-interactions}; Figure~\ref{fig:interaction-mediatype-granularity}). The strongest interaction emerged for perceived ease of use, $F(1, 460) = 21.22$, $p < .001$***, $\eta_p^2 = .044$, with parallel patterns for perceived enjoyment, perceived risk, intention to use, and decision confidence. These interactions indicated that the effect of adding images depended on information granularity.

When information was concrete, image-and-text formats consistently outperformed text-only formats across all five outcomes. For instance, for perceived ease of use, DIDT ($M = 5.64$, $SD = 1.14$) were rated higher than DT ($M = 4.85$, $SD = 1.53$). Simple-effect tests for concrete information yielded small-to-medium user-experience benefits ($d = .32$–.58; see Table~\ref{tab:media-type-simple-effects}). Visual elements appeared to scaffold processing of detailed, relevant travel information, yielding greater enjoyment, lower perceived risk, and stronger intentions to use the format.

\begin{table*}[ht]
  \centering
  \caption{Simple effects of media type within concrete and abstract information conditions on user-experience outcomes. Higher scores indicate more favorable evaluations; for perceived risk, lower scores indicate lower perceived risk.}
  \label{tab:media-type-simple-effects}
  \begin{tabular}{llccccc}
    \toprule
    Info Granularity & Dependent Variable & Image-and-text $M(SD)$ & Text-only $M(SD)$ & Test statistic & $p$ & Effect size \\
    \midrule
    \multirow{5}{*}{Concrete}
      & Ease of Use & 5.64 (1.14) & 4.85 (1.53) & $t(217.91) = 4.46$ & $p < .001$*** & $d = 0.58$ \\
      & Enjoyment & 4.76 (1.60) & 4.06 (1.64) & $t(232) = 3.31$ & $p = .001$** & $d = 0.43$ \\
      & Risk\textsuperscript{a} & 3.10 (1.29) & 3.49 (1.17) & $t(232) = 2.42$ & $p = .016$* & $d = 0.32$ \\
      & Intention to Use & 3.66 (1.24) & 3.08 (1.38) & $t(230.76) = 3.38$ & $p < .001$*** & $d = 0.44$ \\
      & Decision Confidence & 5.31 (1.31) & 4.89 (1.35) & $t(232) = 2.45$ & $p = .015$* & $d = 0.32$ \\
    \midrule
    \multirow{5}{*}{Abstract}
      & Ease of Use & 5.20 (1.30) & 5.51 (1.08) & $t(228) = 1.95$ & $p = .051$ & $d = 0.26$ \\
      & Enjoyment & 4.53 (1.43) & 4.38 (1.27) & $t(228) = 0.84$ & $p = .400$ & $d = 0.11$ \\
      & Risk\textsuperscript{a} & 3.64 (1.30) & 3.42 (1.35) & $t(228) = 1.24$ & $p = .214$ & $d = 0.17$ \\
      & Intention to Use & 2.86 (1.24) & 2.79 (1.23) & $t(228) = 0.43$ & $p = .668$ & $d = 0.06$ \\
      & Decision Confidence & 5.11 (1.46) & 5.30 (1.32) & $t(228) = 1.03$ & $p = .305$ & $d = 0.14$ \\
    \bottomrule
  \end{tabular}
  \vspace{2pt}
\caption*{\normalfont\footnotesize
\parbox{0.97\linewidth}{
\raggedright
\textit{Note}. Image-and-text formats correspond to AIAT (abstract) and DIDT (concrete) conditions; text-only formats correspond to AT (abstract) and DT (concrete) conditions. All tests compare image-and-text to text-only formats within each information-granularity condition. \\
\textsuperscript{a} In this variable, Lower values indicate better outcomes. \\
*$p < .05$, **$p < .01$, ***$p < .001$
}}
\end{table*}

In contrast, when information was abstract, adding abstract images did not reliably change perceived ease of use, enjoyment, decision confidence, intention to use, or perceived risk compared with text-only formats. All corresponding simple-effect tests for abstract information were non-significant (see Table~\ref{tab:media-type-simple-effects}). This pattern suggests that, for high-level overviews, abstract images may create cognitive interference rather than support processing.

It may be particularly noteworthy that the interaction between media type and information granularity showed no significant effects on cognitive load ($p = .135$), perceived usefulness ($p = .231$), or perceived information credibility ($p = .059$). These non-significant results suggest that, while visual elements differentially impact user experience based on information granularity, they do not fundamentally alter cognitive-processing demands, overall utility perceptions, or trustworthiness assessments across abstraction levels.

%% file: Sections/discussion.tex
\section{Discussion}

\begin{figure}[htbp]
    \centering
    \includegraphics[width=\linewidth]{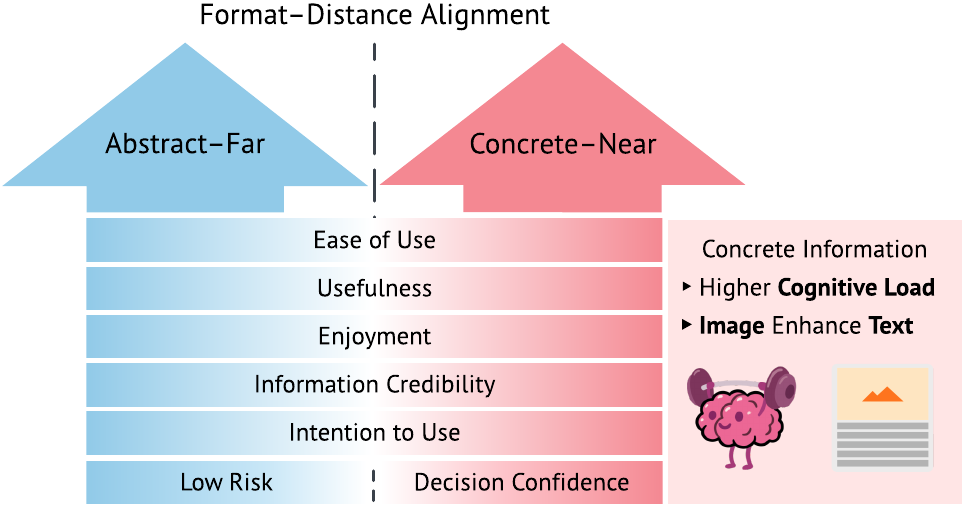}
    \Description{Concept diagram illustrating format and distance alignment. The left side shows Abstract paired with Far distance and the right side shows Concrete paired with Near distance. Six horizontal bands list outcomes: ease of use, usefulness, enjoyment, information credibility, intention to use, and either low risk or high decision confidence. Alignment improves ease of use, usefulness, enjoyment, information credibility, and intention to use. Risk is lower in abstract far contexts while confidence is higher in concrete near contexts. A side box notes that concrete information entails higher cognitive load and that images strengthen concrete text. Decorative icons include a cartoon brain lifting weights and a page with an image above text.}
    \caption{\textbf{Format--distance alignment:} a conceptual synthesis of our empirical findings. Use \textit{abstract} information for \textit{far-distance} planning and \textit{concrete} information for \textit{near-distance} execution. This alignment improves \textit{ease of use}, \textit{usefulness}, \textit{enjoyment}, \textit{information credibility}, and \textit{intention to use}; it also lowers perceived \textit{risk} in abstract–far contexts and raises \textit{decision confidence} in concrete–near contexts. Although concrete formats impose higher \textit{cognitive load}, this additional effort becomes productive when matched to near-distance tasks. Separately, \textit{images} enhance concrete text but provide no reliable benefit for abstract text, indicating that multimedia advantages depend on content-level complementarity.}
    \label{fig:concept}
\end{figure}

The above results demonstrate that aligning IPFs with users' psychological distance from their task produces systematic improvements across cognitive, affective, and behavioral dimensions. They also reveal how CLT and psychological distance provide a robust framework for information design in conversational search systems (see \autoref{fig:concept} for a consolidated design-space view of these effects and mechanisms).

\subsection{Format--Distance Alignment as a New Dimension of Usability}

Various IAM components were rated higher when information-construal level and tasks' psychological distance were aligned. \textit{Perceived ease of use} benefits show how alignment enhances usability by eliminating cognitive translation between representational formats. This extends cognitive-fit theory \cite{vessey1991cognitive} to conversational search contexts, showing that abstract overviews spare far-distance users from unnecessary detail, while concrete specifications eliminate near-distance users' need to extract actionable steps from abstract concepts. \textit{Perceived usefulness} results demonstrated that information value is contextually determined rather than inherent, extending the IAM \cite{sussman2003informational} by showing that identical information can shift from valuable to counterproductive based solely on users' psychological distance. \textit{Perceived information credibility} findings extend AI-trustworthiness research \cite{jacovi2021formalizing}, revealing that beyond source or content quality, credibility also emerges from the pragmatic appropriateness of information presentation: format--distance alignment signals that an information provider understands users' actual needs, enhancing trust by demonstrating contextual awareness \cite{clark1991grounding, grice1975logic}. \textit{Enjoyment} exhibited the largest effect sizes, suggesting that format--distance alignment's benefits extend beyond instrumental ones. Processing-fluency research explains this pattern: ease of processing triggers positive emotional responses and intrinsic pleasure \cite{alter2009uniting, reber2004processing} through the ``hedonic marking'' of fluency \cite{winkielman2001mind}. This extends the affect-as-information framework \cite{schwarz2012feelings}, demonstrating that users' affective responses can signal perceived interaction appropriateness—experiencing pleasure when information formats match their psychological distance.

Decision confidence and perceived risk exhibited curious asymmetric alignment effects: alignment significantly increased confidence only in concrete, near-distance conditions, while significantly reducing perceived risk only in abstract, far-distance conditions. A possible explanation stems from regulatory-focus theory \cite{pennington2003regulatory, higgins1997beyond}, which holds that near psychological distance increases sensitivity to potential losses and negative outcomes \cite{idson2000distinguishing}. In our case, participants performing near-distance tasks could have been attuned to concrete details pertinent to safety, security, and minimizing errors, while being especially sensitive to the absence of such details \cite{idson2000distinguishing}. Consequently, when information lacked concrete details, near-distance participants perceived significantly higher risk than their far-distance counterparts. In contrast, concrete details likely proved particularly valuable in boosting decision confidence among near-distance users, who expected such specifics. For far-distance users, however, excessive details may be constraining \cite{forster2003speed} or undermine confidence by exposing implementation uncertainties, a pattern consistent with the ``planning fallacy'' \cite{buehler2003planning}.

These numerous benefits suggest that format--distance alignment should be adopted as a user-centered design dimension for interactive AI systems. Current systems either assume relatively stable user characteristics discoverable through interaction history \cite{ait2023power, ma2021one}, using them to deliver personalized content (e.g. suggested search results), or employ pre-defined heuristics such as anthropomorphism and linguistic styles. As such, they generally neglect how \textbf{fundamental information structure should adapt to users' shifting construal states, which transcend personal preferences and existing heuristics} \cite{trope2010construal}. Section \ref{discussion:designImplications} discusses concrete design implications.

\subsection{Interactive Effects of Processing Fluency and Cognitive Load}

While concrete information consistently created higher cognitive loads than abstract information, it also created better user outcomes when matched with near psychological distances. This contradicts common wisdom that interface designs should always aim to minimize cognitive load \cite{sweller2011cognitive}.
In part, this could be explained by cognitive-load theory's \cite{sweller2011cognitive} distinction between \textit{intrinsic} load (from inherent task or information complexity) and \textit{germane} load (from resulting knowledge and schema construction) \cite{de2010cognitive}. Construal-congruent processing provides a lens for understanding the link between intrinsic and germane load in a conversational-search context. Prior research suggests that higher cognitive load can create greater germane load under the right conditions \cite{schnotz2007reconsideration}. Therefore, when concrete information is presented together with near-distance tasks, the germane load this induces may ultimately enhance user outcomes such as satisfaction and decision confidence, by prompting mental processes that best help users achieve their desired outcomes. 

Our linguistic analysis further clarifies the nature of the cognitive load observed in our study. Specifically, the elevated use of closure markers in mismatched conditions suggests that the cognitive load experienced there was extraneous and aversive. According to cognitive closure theory \cite{kruglanski1996motivated}, when individuals face poorly structured or cognitively taxing information environments, they are motivated to terminate the processing effort quickly. This manifests as seizing ``on a solution and freezing'' exploration, resulting in the rigid travel plans we observed. This theoretical lens resolves the paradox in our results where mismatched users employed definitive language despite low decision confidence. Their use of closure markers did not signal certainty in their choices, but rather a desire to escape a disfluent interaction by artificially forcing a conclusion. In contrast, the matched conditions fostered what we term \textit{productive uncertainty}. Despite the high intrinsic load of concrete information (H1), the alignment with near-distance needs allowed users to utilize that load effectively (germane load). This is evidenced by their increased use of exploratory language, indicating they remained cognitively open to weighing trade-offs. Thus, format-distance alignment does not merely make tasks ``easier''; it shifts the user's cognitive mode from defensive task termination to engaged, effective deliberation.

A further explanation may lie in the interaction between cognitive load and processing fluency \cite{reber2004processing}. Previous work distinguished between \textit{expected} and \textit{experienced} fluency \cite{jiang2014feels}: i.e., under high cognitive load, users' \textit{a priori} expectations about processing fluency are overridden, and they instead evaluate interactions based on their actual fluency. When concrete information is matched with near-distance tasks, this synergy between desired and provided information can be expected to raise experienced fluency, resulting in better user outcomes (in spite of possibly lower expected fluency, which is less salient under higher cognitive loads). 

In summary, the relationship between cognitive load and user experience depends critically on format--distance alignment rather than absolute processing demands \cite{paas2016cognitive}, contrary to common assumptions that interface design should uniformly minimize the latter \cite{sweller2011cognitive}. This has important implications for the design of informational-retrieval and other digital systems (see section \ref{discussion:designImplications}).

\subsection{Multimedia Formats and the Importance of Information Granularity}

Consistent with past findings that visual elements \cite{mayer2005cambridge, clark1991dual} and multimodal presentation \cite{mayer2002multimedia, paivio1991dual} tend to enhance information processing, we found that across multiple key dimensions of user experience, \textbf{image-and-text formats on aggregate outperformed text-only ones. Interestingly, however, this was only true when the presented information was concrete}. 
One explanation for this comes from theories of cognitive load and channel-capacity constraints \cite{sweller2011cognitive, kalyuga2011cognitive}. Both imagery and textual information are cognitively processed through visual channels \cite{mayer2001cognitive} and thus compete for the same cognitive resource, causing a ``split-attention effect'' \cite{ayres2005split, chandler1991cognitive}. This can cause interference rather than enhancement \cite{schnotz2005integrated}, suggesting that adding images may not always improve user outcomes.

In our case, concrete images of the specific attractions referenced in the corresponding text provided rich perceptual information about ambiance, aesthetic qualities, and spatial configurations that are difficult to represent textually. Therefore, both text and image can be said to have contributed unique, non-redundant information \cite{kozhevnikov2007cognitive, kosslyn1996image}, thus helping these participants construct more comprehensive mental models \cite{barsalou2008grounded, zwaan2004immersed}. Conversely, in abstract image-and-text conditions, images were location maps, which were not necessarily critical to gaining an overall understanding of a travel destination, and thus could have been interesting but irrelevant, impairing information absorption \cite{sundararajan2020keep, rey2012review}. Notably, ``abstract'' images other than maps could also be construed as similarly unhelpful to understanding abstract text. For instance, at the time of writing, a Google search for ``Mexico travel destinations'' yields AI-retrieved wide-shots of different Mexican cities. While aesthetically appealing, these seem unlikely to help a user actually plan a holiday in Mexico. This illustrates that differences in visual specificity and semantic alignment between maps and photographs can shape media-type patterns, alongside information granularity.

Our findings therefore provide additional perspective on a continuing question in multimedia learning research: why visual augmentation sometimes enhances and sometimes impairs information processing \cite{mayer2005cambridge, schnotz2005integrated}. We suggest that \textbf{information granularity serves as a critical moderator} explaining these contradictory findings, while recognizing that other stimulus properties such as visual richness and semantic alignment are also likely to contribute. 

\subsection{Design Implications for Adaptive Information Presentation in Conversational Search Systems}
\label{discussion:designImplications}

Moving beyond previous work that compared conversational and web search based on psychological distance \cite{yang2025understanding}, our results focus on the conversational modality and suggest specific mechanisms through which these systems can adjust their responses to enhance user experience. Below, we propose three concrete design guidelines for adaptive conversational systems: 1) \textbf{adjusting information granularity} by psychological distance, 2) choosing \textbf{media type that complements this granularity}, and 3) \textbf{treating cognitive load as productive investment rather than a burden}. Each is framed as a mapping from observable user cues to response strategies that can be implemented as decision rules or model policies in deployed systems.

\subsubsection{Match Information Granularity to Psychological Distance}
\label{sec:migcl}

A key finding is that conversational AI systems should \textbf{detect users' cognitive states from linguistic cues, and adjust their responses accordingly.}
Such cues may include temporal markers (``next week'' vs ``next year''), spatial markers (locations close to or far from the user), and task-framing markers that separate exploratory from implementation-focused intentions.
Systems may then tailor their responses accordingly, providing concrete specifications or abstract overviews depending on detected distance.
As an example, if a user makes an exploratory query such as ``I am thinking about a trip to Italy next year'', the assistant could respond with an abstract, text-only overview that summarizes typical regions, seasons, and broad budget ranges. If the user instead makes a near-distance, concrete instruction, e.g. ``I need to finalize a three-day itinerary for Rome this weekend'', the same assistant should instead provide a concrete response that enumerates specific attractions, time slots, and ticket prices, and that now adds images to depict key locations. This may also extend beyond single-turn conversations: If a user's queries change in detected psychological distance throughout the conversation, the assistant can constantly shift between concrete and abstract modes and responses accordingly. 
This shows how granularity and media-type choices can be dynamically tied and adapted to inferred psychological distance within a single conversation.

Crucially, this design principle \textbf{transcends the travel planning domain}, and applies to all psychological distance dimensions.
Consider knowledge-seeking tasks as an illustration: With a short query like ``explain black holes'', contextual analysis suggests it might stem from general curiosity; the task then becomes distant in spatial and hypothetical dimensions. Conversely, asking for black hole definitions for an upcoming examination suggests both temporal and hypothetical closeness. Similarly, problem-solving tasks may also inherently imply distances: a request to "debug this code" suggests the problem actually exists (hypothetical proximity) and requires immediate resolution (temporal proximity). Once these distance markers are identified, systems can then tailor their responses accordingly. We acknowledge that a key and distinct challenge here lies in identifying psychological distance cues, presenting a distinct technical challenge. Section \ref{sec:scope} elaborates further on generalizability.

In sum, format--distance alignment represents a distinct and new consideration for user-centered design of conversational AI. When existing design strategies like content personalization are impractical  \cite{toch2012personalization, lam2006you, schein2002methods}, format--distance alignment, based solely on the user's current inputs, provides a powerful and prior-free way to enhance agent design. Further, extending previous research comparing user preferences between conventional web search to conversational search under different psychological distances \cite{yang2025understanding}, this study shows that differing distances can be taken into account entirely within a conversational search system, opening up new ways for optimizing user experiences.

\subsubsection{Optimizing Media Combinations Based on Information Granularity}

As briefly noted above, our finding that images enhanced concrete but not abstract information in a travel-planning context challenges assumptions about multimedia's universal benefits \cite{mayer2005cambridge, clark1991dual}. Conversational AI designers should adopt a granularity-first design principle: i.e., \textbf{ascertain the required abstraction level based on psychological distance, and then select media accordingly while ensuring that available visuals truly complement the associated text}. This reverses current practice, where media decisions tend to precede content structuring and can therefore lead to cognitive interference when misaligned visual and textual elements compete for processing resources \cite{kalyuga2011cognitive, schnotz2005integrated}.

The visual-enhancement patterns emerging from our data reveal important nuances for media selection. When textual travel information was abstract, adding straightforwardly representational images failed to enhance processing because they created redundancy rather than complementarity \cite{sweller2011cognitive}. However, abstract visualizations that encode relationships, structures, or comparisons not easily expressed in text -- such as concept maps, flowcharts, hierarchical diagrams, or data visualizations -- may provide complementary value even for abstract information, by leveraging visual-spatial processing for pattern recognition \cite{larkin1987diagram}. This highlights a key principle of ensuring visual--textual complementarity: \textbf{visual elements should extend rather than duplicate written information}, regardless of abstraction level.

Many contemporary conversational AI systems adopt conservative text-first strategies, likely for reasons including computational costs, response latency, and guarding against inappropriate image generation. Our granularity-based framework suggests a more nuanced approach than uniformly defaulting to text-only or automatically including images. Rather, systems should \textit{selectively} incorporate visual content based on their users' detected abstraction requirements and the complementary value of available imagery. For concrete information queries such as specific product details, procedural instructions, or location-specific requests, photorealistic images and detailed diagrams can provide experiential previews that outshine textual descriptions. For abstract information, on the other hand, systems should prioritize structural visualizations that reveal patterns and relationships while avoiding imagery that is merely decorative and/or redundant with what text already conveys.

\subsubsection{Reframing Load as Investment Rather than Burden}

Our most provocative finding was that increased cognitive load under matched format--distance conditions coincided with enhanced satisfaction. More specifically, the additional mental demands of processing detailed information, when aligned with near psychological distance, functioned as cognitive investment yielding superior decision confidence, rather than extraneous burden degrading user experience. This suggests that system designers should \textbf{distinguish more carefully between productive processing that deepens engagement and wasteful effort that impedes task completion}.

Consequently, we recommend that interfaces \textbf{explicitly frame cognitive effort as valuable when presenting matched information}, as a means of helping their users understand why certain responses demand deeper processing. When delivering detailed specifications for near-distance travel decisions, systems could preface such information with contextualizing statements, such as: ``These details are retrieved directly from booking sites; you may use these to finalize your travel plan.'' Such a framing could transform anticipated cognitive load from a deterrent into a motivator.

This reframing may be particularly useful to conversational AI in its ongoing competition with traditional search interfaces. That is, users accustomed to scanning multiple search results might initially resist providing conversational responses with the focused attention they demand. But by explicitly connecting cognitive investment to decision quality, systems can shift user perception from ``\textit{this is taking too long}'' to ``\textit{this depth helps me decide}''. The key lies not in minimizing all cognitive load, a futile goal in the case of complex decisions that inherently require mental effort, but in ensuring that the load serves users' goals rather than systems' limitations.

\subsection{Limitations and Future Directions}

Our study provides foundational evidence for the importance of format--distance alignment in conversational searches within specific limits. These are considered below.

\subsubsection{Scope and Generalizability}
\label{sec:scope}

First, regarding \textbf{task generalizability}, our study used a travel planning task for methodological reasons: it allowed for explicit temporal and spatial cues, which enabled highly controllable manipulation of psychological distance and helped establish causal relationships between format preferences and psychological distance.
The theoretical foundation of our design principle, however, supports its application across broader domains. Psychological distance is a core metacognitive feature of human judgment \cite{trope2010construal}, and the correspondence between task construal levels (from concrete to abstract) and psychological distance has been consistently demonstrated across many domains, from planning a yard sale or moving house \cite{liberman2002effect} to advertising \cite{choi2019text}, research recruitment \cite{trope2007construal}, and test-taking \cite{nussbaum2006predicting}. In various information-seeking contexts such as healthcare decisions, financial planning, and educational queries, specific cognitive constraints, risk profiles, and information requirements also exist, influencing how psychological distance cues are identified and weighted. As discussed in Section \ref{sec:migcl}, these domain-specific differences primarily constitute technical challenges at the implementation level rather than theoretical limitations; nonetheless, future work must rigorously empirically validate this perspective across a wider range of task contexts. Future research should focus on developing domain-specific cue identification frameworks and establishing systematic methods to capture and evaluate psychological distance indicators across various application contexts, thereby contextualizing this theoretical principle into actionable design guidelines tailored to different scenarios.

Second, regarding \textbf{psychological distance dimensions}, our study manipulated only two of the four dimensions for practical reasons, as including all four would have greatly increased experimental complexity. However, the principle of psychological equivalence \cite{trope2010construal, maglio2013distance} holds that all distance dimensions share a "common meaning" \cite{trope2010construal, bar2007automatic, stephan2010politeness}, follow the same cognitive processes, and similarly influence information needs. Our results support this assumption, showing no significant interaction effects between dimensions \footnote{A full analysis of dimensional interactions is provided in Appendix A.1}. This suggests that, theoretically, our findings generalize from spatial and temporal distance to social and hypothetical distance as well. Nonetheless, future work could empirically validate this theoretical extension and then apply more comprehensive distance manipulations to examine the relative weights and potential dominance of different dimensions in real-world multi-cue settings.

Finally, our participant pool of 464 CloudResearch participants, while diverse in age and gender, consisted primarily of English-speaking adults from the U.S.. Format preferences and psychological-distance perceptions may vary across cultures, particularly given documented differences in holistic versus analytic cognitive styles \cite{nisbett2010geography}.

\subsubsection{Methodological Constraints}

First, our single-turn interaction design aimed to isolate participants' perception and evaluation of response formats. By limiting interactions to single query–response exchanges, we reduced confounds such as conversational flow, social cues, and system adaptivity, allowing effects to be more clearly attributed to format--distance alignment. However, real conversations are dynamic and continuous: users ask follow-up questions, adjust constraints, and request reformulations. In multi-turn contexts, psychological distance and format preferences may shift over the course of the dialogue. Thus, the effects reported here reflect users' immediate reactions to initial outputs rather than their evaluations of full conversational exchanges. Future work should examine the persistence, evolution, and boundary conditions of format--distance alignment effects in multi-turn interactions, including whether the benefits of alignment persist, intensify, or diminish when systems can dynamically adjust response formats based on user feedback.

Second, our two conditions provided different numbers and types of images: the concrete conditions presented three photographs, while the abstract condition provided one. We did this to align the quantity of information provided through text and image. Because concrete texts inherently provide more detail than abstract texts, they naturally call for more images to support that detail. In our study, the concrete texts included three distinct pieces of information (history, cuisine, natural scenery), each with a corresponding image; omitting one or more images would be unintuitive, since some specified information would lack visual support. On the other hand, increasing the number of images in abstract conditions may also be unhelpful, because these images generally provide less useful information. Following CLT, we used overview maps for abstract conditions and content-relevant photos for concrete conditions to align visual and textual abstraction levels, and manipulation checks confirmed that this worked. However, this design means the interaction between media type and granularity may partly reflect differences in visual richness, readability, and perceived relevance, not modality alone. Future work should replicate our results with visual stimuli standardized in resolution and semantic relatedness across conditions, and systematically examine how to design abstract images that effectively complement abstract text.

Finally, our study used a multi-method approach, triangulating self-reported user perceptions with linguistic indicators from written travel plans. This yields robust evidence about users' cognitive states, but our reliance on a subjective preference task limits conclusions about objective decision quality, as we could not test whether plans were empirically optimal or factually correct. Future work should apply this paradigm to high-stakes domains with objective ground truths (e.g., healthcare diagnosis, financial investment) to test whether format-distance alignment improves decisions, or use physiological measures (e.g., eye-tracking, EEG) to directly validate the neural correlates of processing fluency and cognitive load.

\subsubsection{Implementation Feasibility}

Translating our findings into real-world systems will introduce new technical hurdles. Beyond detecting temporal, spatial, and task-framing markers from language, we posit a plausible behavioral pattern that could guide engineering choices: when users are in a near-distance state (low construal), their prompts and queries are likely to be detail-rich, making it easier for systems to extract psychological-distance features. By contrast, far-distance states (high construal) may yield shorter or more concise requests (e.g., ``Thailand travel destinations''), which lack clear markers of users' target task distance. In practice, systems might treat such high-level queries as provisional signals of far distance and, leveraging the advantages of dialogue, proactively elicit missing specifics to calibrate the abstraction level of subsequent responses. The reliability of these hypotheses, their boundary conditions, and their implications for user experience require targeted empirical evaluation and ablation studies in future work.


%% file: Sections/conclusion.tex
\section{Conclusion}

This paper empirically validates the novel idea that aligning IPFs with users' psychological distance is a particularly promising new direction in conversational-search design. Grounded in CLT, our study establishes format--distance alignment as a distinct design dimension; more specifically, it shows that in terms of multiple user-experience outcomes, abstract formats benefit distant planning while concrete ones benefit proximate tasks, regardless of differing cognitive loads.

Looking forward, conversational AI systems should: 1) detect psychological-distance cues from queries and interaction contexts through temporal markers, spatial indicators, and task framing; 2) calibrate response granularity (abstract vs. concrete) and media choice (text-only vs. image-and-text) based on the detected distance; and 3) dynamically reassess distance across multi-turn conversations as users refine their goals. Implementing these principles, by shifting design focus from uniform information formatting to cognitive alignment, should enable conversational agents to adapt to their users' construal levels rather than merely their stated preferences or historical patterns.